\documentclass[preprint2,twocolumn,times,tighten]{aastex63}

\usepackage{color}
\usepackage{enumitem}
\usepackage{hyperref}
\usepackage{verbatim}
\usepackage{amsmath,amstext,mathrsfs}
\usepackage[all]{hypcap} 
\usepackage{afterpage}    
\usepackage{marginnote}
\usepackage{makecell}
\usepackage{subfigure}
\usepackage{multirow}
\usepackage{lineno}
\shorttitle{Probing 3D Magnetic Fields using PPV cube}
\shortauthors{Hu, Lazarian \& Xu}
\graphicspath{{./}{figures/}}
\begin{document}

\title{Anisotropic Turbulence in Position–Position–Velocity Space: Probing Three-Dimensional Magnetic Fields}

\email{yue.hu@wisc.edu; alazarian@facstaff.wisc.edu; sxu@ias.edu;}

\author[0000-0002-8455-0805]{Yue Hu}
\affiliation{Department of Physics, University of Wisconsin-Madison, Madison, WI 53706, USA}
\affiliation{Department of Astronomy, University of Wisconsin-Madison, Madison, WI 53706, USA}
\author{A. Lazarian}
\affiliation{Department of Astronomy, University of Wisconsin-Madison, Madison, WI 53706, USA}
\author[0000-0002-0458-7828]{Siyao Xu}
\affiliation{Institute for Advanced Study, 1 Einstein Drive, Princeton, NJ 08540, USA\footnote{Hubble Fellow}}

\begin{abstract}
Direct measurements of three-dimensional magnetic fields in the interstellar medium (ISM) are not achievable. However, the anisotropic nature of magnetohydrodynamic (MHD) turbulence provides a novel way of tracing the magnetic fields. Guided by the advanced understanding of turbulence's anisotropy in the Position–Position–Velocity (PPV) space, we extend the Structure-Function Analysis (SFA) to measure both the three-dimensional magnetic field orientation and Alfv\'{e}n Mach number $\rm M_A$, which provides the information on magnetic field strength. Following the theoretical framework developed in \citet{2016MNRAS.461.1227K}, we find that the anisotropy in a given velocity channel is affected by the inclination angle between the 3D magnetic field direction and the line-of-sight as well as media magnetization. We analyze the synthetic PPV cubes generated by incompressible and compressible MHD simulations. We confirm that the PPV channel's intensity fluctuations measured in various position angles reveal plane-of-the-sky magnetic field orientation. We show that by varying the channel width, the anisotropies of the intensity fluctuations in PPV space can be used to simultaneously estimate both magnetic field inclination angle and strength of total magnetic fields.
\end{abstract}

\keywords{Interstellar medium (847); Interstellar magnetic fields (845); Interstellar dynamics (839)}

\section{Introduction}
\label{sec:intro}

Magnetic fields and turbulence are essential in astrophysical studies \citep{1981MNRAS.194..809L,1995ApJ...443..209A,2010ApJ...710..853C,2012ARAA...50..29,Han17,Hu20,cluster,2020A&A...641A..12P}. In the past decades, several methods have been proposed to probe the magnetic fields in various length scales and multiple interstellar phases. In a dusty interstellar medium (ISM), the aligned dust grains polarize the starlight \citep{2000AJ....119..923H,2015MNRAS.452..715P} and dust thermal emissions so that they can reveal the plane-of-the-sky (POS) component of magnetic fields \citep{2007JQSRT.106..225L,2007ApJ...669L..77L,Aetal15}. Synchrotron emission provides another probability to trace the POS magnetic fields in the hot gas-filled region \citep{2006AJ....131.2900C,2016A&A...596A.103P,LP16}. As for the line-of-sight (LOS) signed magnetic field strength, Zeeman splitting \citep{2004mim..proc..123C,2012ARAA...50..29} and Faraday rotation \citep{1996ApJ...458..194M,2006ApJS..167..230H,2015A&A...575A.118O,2016ApJ...824..113X} are commonly used. Recently a number of new methods can also give novel insight, including the Goldreich-Kylafis effect \citep{1981ApJ...243L..75G,1982ApJ...253..606G}, the
atomic/ionic ground state alignment (GSA) effect \citep{2006ApJ...653.1292Y,2007ApJ...657..618Y,2008ApJ...677.1401Y,2012JQSRT.113.1409Y}, the line-width difference in weakly ionized medium \citep{2008ApJ...677.1151L,2015ApJ...810...44X}. Nevertheless, these measurements can only trace two-dimensional magnetic fields (either the LOS or POS component) in either two-dimensional plane or three-dimensional space. 

Unlike the in-situ measurements in the solar wind, probing three-dimensional magnetic fields, including both the LOS or POS components, in three-dimensional space is notoriously challenging. Based on MHD turbulence's anisotropic properties, \citet{LY18b} proposes to trace the three-dimensional magnetic fields through synchrotron polarization. This method utilizes multiple-wavelength measurements of synchrotron polarization. By taking gradients of the wavelength derivative of synchrotron polarization, one can recover the direction of three-dimensional magnetic fields, i.e., obtaining the POS and LOS components simultaneously. Later, \citet{2019MNRAS.485.3499C} showed that dust thermal emission's depolarization properties contain the information of the LOS magnetic fields. Consequently, the three-dimensional magnetic fields on a molecular cloud scale can be revealed from the observed polarization fraction.

In addition to the two methods, \citet{SFA} recently propose to trace the three-dimensional magnetic fields through the second-order structure-function of velocity fluctuations, which is termed as the Structure-Function Analysis (SFA; \citealt{SFA,XH21}). The SFA is rooted in MHD turbulence theory \citep{GS95} and turbulent reconnection theory \citep{LV99,2020PhPl...27a2305L}, which revealed the anisotropy of MHD turbulence. As the amplitude of velocity fluctuations is anisotropic, at the same separation away from the eddy's center, the maximum amplitude appears in the direction perpendicular to the local magnetic fields while the minimum amplitude indicates the magnetic field direction \citep{2000ApJ...539..273C,2001ApJ...554.1175M,2002ApJ...564..291C,2003MNRAS.345..325C}. By measuring the minimum velocity fluctuations in the real space, SFA could simultaneously reveal the three-dimensional magnetic field orientation and strength. This measurement requires velocity and three-dimensional distance information, which is only achievable by the GAIA's young star survey \citep{2016A&A...595A...1G,2018A&A...616A...1G,Ha20}. 

For this study, we aim at improving the SFA to be a more general approach in tracing the three-dimensional magnetic field and enabling the usage of Position-Position-Velocity (PPV) cubes. Extracting velocity information from PPV space is non-trivial \citep{LP00}. One of the most common ways is using the velocity centroid, i.e., the moment-one map. Earlier studies of velocity centroids have been developed to reveal the direction of the POS magnetic field through the structure-function \citep{2002ASPC..276..182L,2005ApJ...631..320E,2011ApJ...740..117E,2014ApJ...790..130B,2017MNRAS.464.3617K,XH21}. In this work, instead of using the velocity centroid, we explore the second way of extracting velocity fluctuations from the velocity caustic effect in PPV cubes. The concept of velocity caustic was firstly explained by \citet{LP00}. It reveals that the observed intensity distribution in a PPV channel is regulated by turbulent velocity and thermal velocity along the LOS. \citet{2016MNRAS.461.1227K} analytically implemented the structure-function to velocity channels. They found the anisotropy of observed intensity distribution has a dependence on the channel width. Based on this finding, we further elaborate the anisotropy at a given channel width is related to Alfv\'en Mach number $\rm M_A$ and the inclination angle $\gamma$ of the 3D magnetic fields and the LOS. We combine these insights, along with the SFA, to trace three-dimensional magnetic fields and magnetization $\rm M_A^{-1}$ using PPV cubes. 

The paper is organized as follows. In \S~\ref{sec:theory}, we introduce the theory of MHD turbulence and analyze the anisotropy of the velocity structure functions in PPV channels. In \S~\ref{sec.method}, we illustrate the recipe of SFA to trace three-dimensional magnetic fields. In \S~\ref{sec:data}, we provide the details of the incompressible and compressible MHD simulations used in this work.  In \S~\ref{sec:results}, we perform numerical experiments to test the new technique SFA. We present discussion in \S~\ref{sec.dis} and summary in \S~\ref{sec:con}, respectively.

%%%%%%%%%%%%%%%%%%%%%%%%%%%%%%%%%%%%%%%%%%%%%%%%%%%%%
\section{Theoretical formulation of the SFA}
\label{sec:theory}
\subsection{Anisotropy of MHD turbulence}
\begin{figure*}
\centering
\includegraphics[width=1.0\linewidth,height=0.67\linewidth]{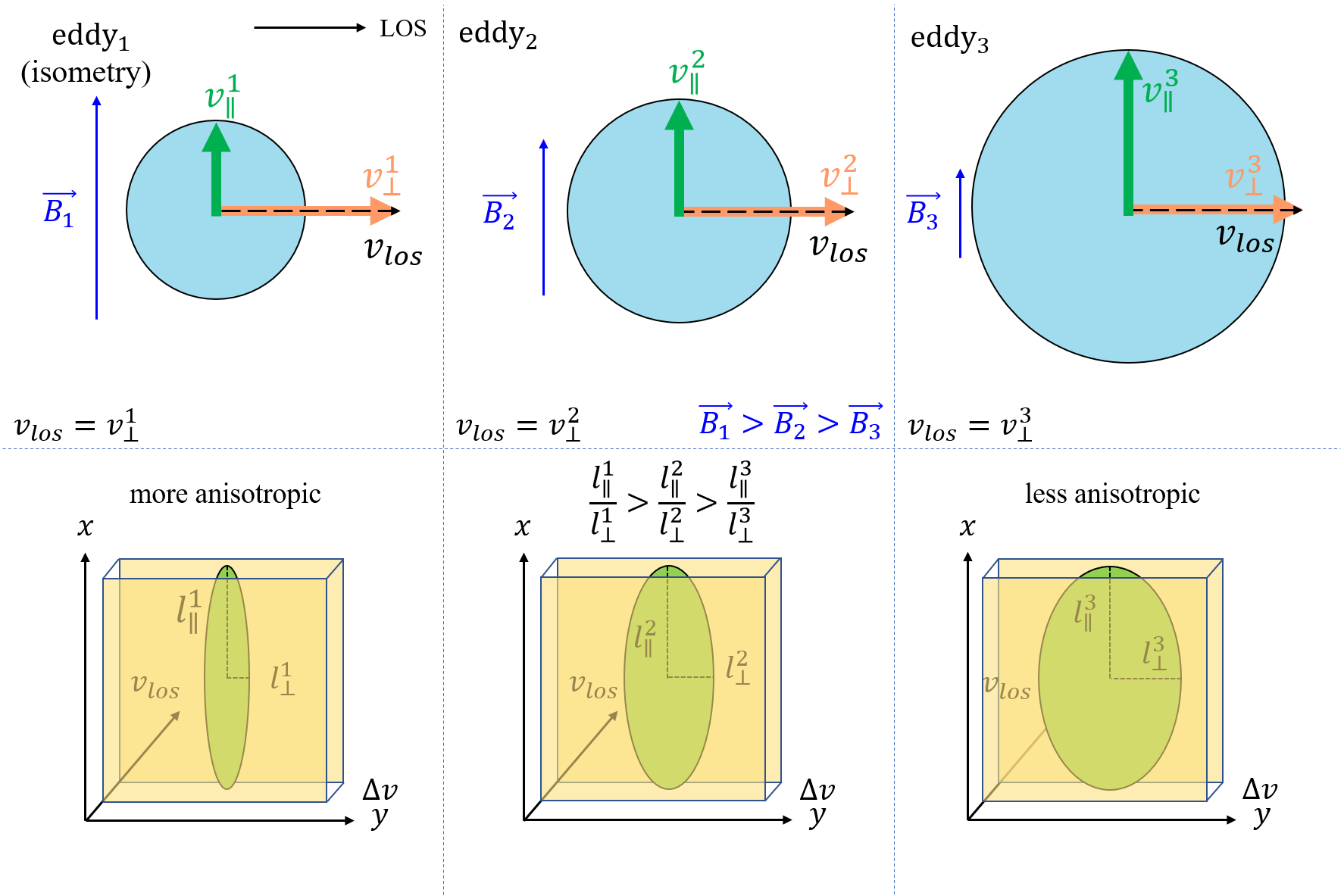}
\caption{\label{fig:Ma} An illustration of how magnetic field strength affects eddies' mapping from real PPP space to PPV space. Three isometric eddies (eddy$_1$, eddy$_2$, and eddy$_3$) have different magnetic fields ($\vec{B}_1>\vec{B}_2>\vec{B}_3$) which are perpendicular to the LOS in PPP space. The amplitude of velocity fluctuations for isometric eddy is anisotropic, i.e., the maximum amplitude $v_\bot$ appears in the direction perpendicular to the local magnetic fields. In contrast, the minimum amplitude $v_\parallel$ is in the parallel direction. The LOS velocity $v_{los}$ only consists of the turbulent velocity $v_\bot$, which is perpendicular to the magnetic field. For a given amplitude of $v_{los}=v_\bot^1=v_\bot^2=v_\bot^3$, strong magnetic field induces more significant anisotropy (i.e., $v_\parallel^1<v_\parallel^2<v_\parallel^3$, see Eq.~\ref{eq.vv}). Three eddies (in real PPP space; top panel) are being mapped to the PPV space (bottom panel) with identical channel width $\Delta v$ (yellow box). The observed intensity fluctuation corresponding to eddy$_1$'s case is more anisotropic ($l_\parallel^1/l_\bot^1>l_\parallel^2/l_\bot^2>l_\parallel^3/l_\bot^3$).}
\end{figure*}
 
The developement of MHD turbulence theory is a long story. The earliest model of MHD turbulence is isotropic \citep{1963AZh....40..742I,1965PhFl....8.1385K}. Later, this model was revised by a number of theoretical and numerical studies revealing that the MHD turbulence is anisotropic in sub-Alfv\'{e}nic condition and is isotropic in large-scale sup-Alfv\'{e}nic condition \citep{1981PhFl...24..825M,1983JPlPh..29..525S,1984ApJ...285..109H,1995ApJ...447..706M}. Thereafter, a key concept of "critical balance" condition, i.e., equating the cascading time ($k_\bot v_l$)$^{-1}$ and the wave periods ($k_\parallel v_A$)$^{-1}$, was given by  \citet{GS95}, denoted as GS95. Here $k_\parallel$ and $k_\bot$ are the components of the wavevector parallel and perpendicular to the magnetic field, respectively. $v_l$ is turbulent velocity at scale $l$ and $v_A$ is Alfv\'{e}n speed. By considering the Kolmogorov-type turbuelence, i.e., $v_l\propto l^{1/3}$, the GS95 anisotropy scaling be can easily obtained: 
\begin{equation}
    k_\parallel\propto k_\bot^{2/3}
\end{equation}
which reveals the anisotropic nature of turbulence eddies, i.e., the eddies are elongating along the magnetic fields. Nevertheless, the GS95's consideration is drawn in the global reference frame, i.e., the direction of wavevectors is defined with respect to the mean magnetic field. Due to the averaging effect along the LOS, only the largest eddy are dominant in the global frame so that the observed anisotropy appears to be scale-independent \citep{2000ApJ...539..273C}.

The scale-dependent anisotropy is later obtained from the study of fast turbulent reconnection (\citealt{LV99}, denoted as LV99), which considered a local reference frame. The local reference frame is defined with respect to the magnetic field passing through the eddy at scale $l$. LV99 explained that only the motion of eddies perpendicular to the local magnetic field direction obeys the Kolmogorov law ( i.e., $v_{l,\bot}\propto l_\bot^{1/3}$ ), because the magnetic field gives minimal resistance along this direction. Considering the "critical balance" in the local reference frame: $v_{l,\bot}l_\bot^{-1}\approx v_Al_\parallel^{-1}$, the scale-dependent anisotropy scaling is then:
\begin{equation}
\label{eq.lv99}
 l_\parallel= L_{inj}(\frac{l_\bot}{L_{inj}})^{\frac{2}{3}}{\rm M_A^{-4/3}}, {\rm M_A\le 1}
\end{equation}
where $l_\bot$ and $l_\|$ are the perpendicular and parallel scales of eddies with respect to the {\it local} magnetic field, respectively. $L_{inj}$ is the turbulence injection scale. The corresponding anisotropy scaling for velocity fluctuation is:
\begin{equation}
\label{eq.v}
\begin{aligned}
 v_{l,\bot}&= v_{inj}(\frac{l_\bot}{L_{inj}})^{\frac{1}{3}}{\rm M_A^{1/3}}\\
 &= v_{inj}(\frac{l_\parallel}{L_{inj}})^{\frac{1}{2}}{\rm M_A}, {\rm M_A\le 1}
\end{aligned}
\end{equation}
where $v_{inj}$ is the injection velocity. This scale-dependent anisotropy in the local reference frame was numerically demonstrated
\citep{2000ApJ...539..273C,2001ApJ...554.1175M} and in-situ observations in solar wind \citep{2016ApJ...816...15W}.

\subsection{Nature of PPV space}
The observed intensity distribution in a given spectral line in PPV space is defined by both of emitters' density and their velocity distribution along the LOS. The LOS component of velocity $v_{los}$ at the position $(x,y)$ in the POS is a sum of the turbulent velocity $v(\Vec{r})$ and the residual component due to thermal motions. This residual thermal velocity $v_{los}-v(\Vec{r})$ has a Maxwellian distribution so that the gas distribution $\rho(x,y,v)$ in PPV cubes and $\rho(\Vec{r})$ in real-space has the relation \citep{LP04}:
\begin{equation}
\label{eq.max}
\rho(x,y,v_{los})=\int_{-S}^S\frac{\rho(\Vec{r})}{\sqrt{2\pi\beta_T}}\exp[-\frac{(v_{los}-v(\Vec{r}))^2}{2\beta_T^2}]dz
\end{equation}
where $\beta_T=k_BT/m$, $m$ being the mass of atoms, $k_B$ is the Boltzmann constant, and $T$ being the temperature, which can vary from point to point if the emitter is not isothermal.

\subsubsection{Dependence on $\rm M_A$}
%When mapping the turbulent eddies from PPP space to PPV space, the maximum amplitude of turbulent velocity is determined by channel width $\Delta v$. We also know that small eddies exhibit more significant anisotropy. Therefore, it is natural to explore how magnetic fields affect the observed anisotropy in a given channel.
\begin{figure*}
\centering
\includegraphics[width=1.0\linewidth,height=0.67\linewidth]{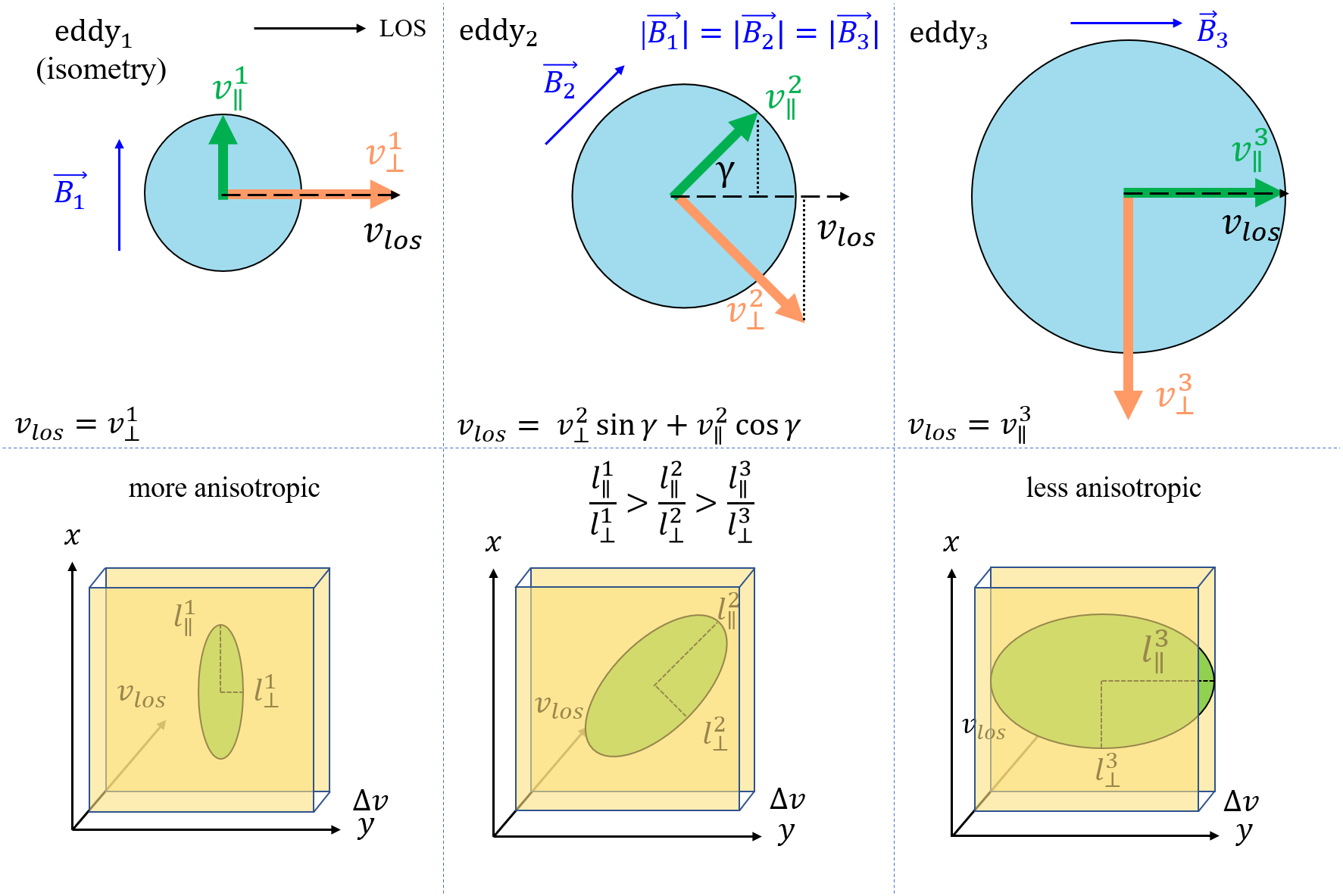}
\caption{\label{fig:gamma} An illustration of how inclination angle affects the mapping of eddies from the real space to the PPV space. Three isometric eddies (eddy$_1$, eddy$_2$, and eddy$_3$) have identical magnetic fields strength ($|\vec{B}_1|=|\vec{B}_2|=|\vec{B}_3|$). $\vec{B}_1$ is perpendicular to the LOS ($\gamma=\pi/2$), $\vec{B}_2$ is inclined to the LOS with angle $\gamma$, and $\vec{B}_3$ is parallel to the LOS ($\gamma=0$). The LOS velocity $v_{los}=v_\bot\sin\gamma+v_\parallel\cos\gamma$, in which $v_\bot$ and $v_\parallel$ are components of turbulent velocity perpendicular and parallel to the magnetic field, respectively. For a given amplitude of $v_{los}$, eddy$_1$ (i.e., $\gamma=\pi/2$) is more  anisotropic, as $v_\bot>v_\parallel$ (see Eq.~\ref{eq.vv}) and  $v_\parallel^1/v_\bot^1>v_\parallel^3/v_\bot^3>v_\parallel^3/v_\bot^3$. Three eddies (in real PPP space; top panel) are being mapped to the PPV space (bottom panel) with identical channel width $\Delta v$ (yellow box). The observed intensity fluctuation corresponding to eddy$_1$'s case is more anisotropic ($l_\parallel^1/l_\bot^1>l_\parallel^2/l_\bot^2>l_\parallel^3/l_\bot^3$). Different from Fig.~\ref{fig:Ma}, this difference in anisotropy is induced by inclination angle. }
\end{figure*}

In Fig.~\ref{fig:Ma}, we consider a simple case that the magnetic fields are perpendicular to the LOS. There three isometric eddies (eddy$_1$, eddy$_2$, and eddy$_3$ in PPP space) have different magnetic fields ($\vec{B}_1>\vec{B}_2>\vec{B}_3$). For a given channel width $\Delta v$, for example $\Delta v= 1$ km s$^{-1}$, the amplitude of maximum LOS velocity $v_{los}$ is then determined. However, here $v_{los}$ is only contributed by the perpendicular component of turbulent velocity. The studies in \citet{SFA} and \citet{XH21} derived that the amplitude of velocity fluctuations for isometric eddy is anisotropic, i.e., the maximum amplitude $v_\bot$ appears in the direction perpendicular to the local magnetic fields while the minimum amplitude $v_\parallel$ is in the parallel direction. In particular, the anisotropy of turbulent velocity (i.e., the ratio of $v_\bot^2/v_\parallel^2$) is a power-law relation with $\rm M_A$:
\begin{equation}
\label{eq.vv}
  v_\bot^2/v_\parallel^2=
\begin{cases}
       (l_\parallel/L_{inj})^{-1/3}{\rm M_A}^{-4/3}, &{\rm (local, M_A\le1)}\\
       %(l_\parallel/l_{a})^{-1/3}{\rm M_A}^{-1/3}, &{\rm (local, M_A>1)}\\
        {\rm M_A}^{-4/3}, &{\rm (global, M_A\le1)}\\
\end{cases}
\end{equation}
Here "local" means the measurement is performed in the local reference frame, i.e., the parallel and perpendicular directions are defined by local magnetic fields. "global" means global mean magnetic fields define the global reference frame, i.e., the parallel and perpendicular directions.

In our case, the LOS velocity $v_{los}$ only consists of the turbulent velocity $v_\bot$, which is perpendicular to the magnetic field. For a given channel width, we have $v_{los}=v_\bot^1=v_\bot^2=v_\bot^3$. Consequently, a strong magnetic field ($\vec{B}_1>\vec{B}_2>\vec{B}_3$) corresponds to a small value of $v_\parallel$ (i.e., $v_\parallel^1<v_\parallel^2<v_\parallel^3$, see Eq.~\ref{eq.vv}). It indicates a more anisotropic turbulent velocity field $v(\Vec{r})$. These three isotrometic eddies (in real PPP space) are then being mapped to the PPV space with identical channel width $\Delta v$. As shown in Eq.~\ref{eq.max}, the observed intensity in a PPV channel is related to the distribution of turbulent velocity $v(\Vec{r})$\footnote{Note that the intensity fluctuation in a thin velocity channel is dominated by velocity fluctuations \citep{LP00}.}. An anisotropic $v(\Vec{r})$ would result in an anisotropic intensity distribution \citep{LP00}. Therefore, the observed intensity is more anisotropic for a strongly magnetized medium.
\begin{table*}
\centering
\label{tab:param}
\begin{tabular}{| c | c |}
\hline
Parameter & Meaning \\\hline\hline
$\vec{r}$ & 3D separation \\
$\vec{R}$ & 2D sky separation \\
$\phi$ & 2D angle between $\vec{R}$ and the POS magnetic field \\
$\theta$ & Angle between LOS $\hat{z}$ and $\vec{r}$\\
$\gamma$ & Angle between LOS and mean magnetic field direction\\
$\mu$ & Cosine of the angle between $\vec{r}$ and mean magnetic field direction\\
$\Delta v$ & Channel width\\
$\xi_I(R,\phi,\Delta v)$ & Intensity correlation function\\
$D(R,\phi,\Delta v)$ & Intensity structure function\\
$D_z(\vec{r})$ & z-projection of velocity structure function\\
$W(v)$ & Window function\\
$\beta_T$ & Thermal velocity\\
$\rm M_A$ & Alfv\'en Mach number\\
$\rm M_S$ & Sonic Mach number\\
$v_{los}$ & LOS velocity\\
$v(\vec{r})$ & Turbulent velocity\\
$v_\bot$ & Velocity fluctuation perpendicular to local magnetic field\\
$v_\parallel$ & Velocity fluctuation parallel to local magnetic field\\
$\epsilon$ & Emissivity\\
$\bar{\rho}$ & Mean density\\
\hline
\end{tabular}
\caption{List of notations used in this paper}
\end{table*}

\subsection{Dependence on inclination angle of magnetic fields}
In a real scenario, the isometric eddies may incline to the LOS with angle $\gamma$. This inclination changes the observed anisotropy in the PPV channel. In Fig.~\ref{fig:gamma}, we consider three magnetized isometric eddies (eddy$_1$, eddy$_2$, and eddy$_3$). The magnetic field strength are identical ($|\vec{B}_1|=|\vec{B}_2|=|\vec{B}_3|$), but $\vec{B}_1$ is perpendicular to the LOS, $\vec{B}_2$ inclines to the LOS with angle $\gamma$, and $\vec{B}_3$ is parallel to the LOS. We consider mapping the eddies to a given channel width $\Delta v$. For eddy$_1$, $v_{los}$ is purely contributed by $v_\bot$. However, $v_{los}$ of eddy$_2$ consists of both $v_\bot$ and $v_\parallel$. Considering the fact that $v_\parallel<v_\bot$ and the $v_\bot\propto l_\bot^{1/3}$, the only way to achieve the same amplitude of $v_{los}$ is to increase the eddy's size. Eddy$_3$ is facing a similar situation. As $v_{los}$ of eddy$_3$ only comes from $v_\parallel$, the eddy must further increase its size to get the same $v_{los}$ value. However, this increment of size results in a smaller anisotropy of turbulent velocity, i.e., a smaller ratio of $v_\bot^2/v_\parallel^2$ (see Eq.~\ref{eq.vv}), as well as the observed intensity. This change of anisotropy induced by inclination angle is distinguishable from the one induced by magnetic field strength. It is clear that eddy$_3$'s anisotropy (i.e., $\gamma=0$ case) is not observable in PPV space, while it is not the case when only magnetic field strength gets changed (i.e., $\gamma\ne0$).

The anisotropy of observed intensity in a PPV channel is thus constrained by both $\rm M_A$ and $\gamma$. By measuring two PPV channels' anisotropies, at least $\rm M_A$ and $\gamma$ could be determined simultaneously. In the following, we will analytically show the anisotropy's dependence on $\rm M_A$ and $\gamma$ through the second-order structure-function.

\subsection{Structure function analysis}
The statistical description for a turbulent field within PPV space is firstly performed by \citet{LP00}. Later, \citet{LP04} derived that for optically thin lines the intensity correlation function $\xi_I(R,\phi,\Delta v)$ is \footnote{Note that here the measurement can only be performed in the global reference frame.}:
\begin{equation}
\begin{aligned}
\label{eq.cf}
      \xi_I(R,\phi,\Delta v)\propto\frac{\epsilon^2\bar{\rho}}{2\pi}\int_{-S}^{S}dz[1+\widetilde{\xi}_\rho(\vec{r})][D_z(\vec{r})+2\beta_T]^{-1/2}\\
      \times\int_{-\Delta v/2}^{+\Delta v /2}dv_{los} W(v_{los})\exp[-\frac{v_{los}^2}{2(D_z(\vec{r})+2\beta_T)}]
\end{aligned}
\end{equation}
where $R$ is the 2D separation of two points in the POS, $\phi$ is the position angle that $R$ makes with the POS magnetic field, and $\Delta v$ gives the channel width. $\epsilon$ is emissivity, $\bar{\rho}$ is the mean density, S is the LOS distance. The integration along the LOS involves the overdensity correlation $\widetilde{\xi}_\rho(\vec{r})=\bar{\rho}^2(r_0/r)^\nu$, the projected (along the LOS) velocity structure function $D_z(\vec{r})$, the thermal broadening term $\beta_T=k_BT/m$, where $T$ is temperature, $k_B$ is Boltzmann constant, and $m$ is molecule/atoms' mass . $\vec{r}=\vec{r}_1-\vec{r}_2$ means the 3D separation of two points. The integration over velocities is described by the window function $W(v_{los})$. In terms of Alfv\'en waves, as they are incompressible and do not create any density fluctuations, the overdensity correlation $\widetilde{\xi}_\rho(\vec{r})$ must be zero.

Compared with the intensity correlation function, the intensity structure function $D(R,\phi,\Delta v)$ is better at describing intensity statistics at small scales \citep{LP04}. $D(R,\phi,\Delta v)$ can be expressed as:
\begin{equation}
\begin{aligned}
D(R,\phi,\Delta v)=2[\xi_I(0,0,0)-\xi_I(R,\phi,\Delta v)]
\end{aligned}
\end{equation}
The isotropy degree is defined as;
\begin{equation}
\begin{aligned}
    {\rm iso(R,\gamma, M_A, \Delta v)}&=\frac{D(R,0,\Delta v)}{D(R,\pi/2,\Delta v)}\\
    &=\frac{\xi_I(0,0,0)-\xi_I(R,0,\Delta v)}{\xi_I(0,0,0)-\xi_I(R,\pi/2,\Delta v)}
\end{aligned}
\end{equation}
here $\gamma$ is the inclination angle of 3D magnetic field with respect to the LOS and $\rm M_A$ is the Alfv\'en Mach number. $\gamma$ and $\rm M_A$ are implicitly included in $D(R,\phi,\Delta v)$ (see Appendix~\ref{appendix.A}). The dependence of ${\rm iso(R,\gamma, M_A, \Delta v)}$ will be shown in the following.

The analytical calculation of ${\rm iso(R,\gamma, M_A, \Delta v)}$ was carried by \citet{2016MNRAS.461.1227K}. Here we just briefly summarize it. Computing Eq.~\eqref{eq.cf} requires the knowledge of $D_z(\vec{r})$, which can be obtained from the projection of structure function tensor for the velocity field:
\begin{equation}
\label{eq.9}
    \begin{aligned}
      D_z(\vec{r})=&\langle(v_i(\vec{r}_1)-v_i(\vec{r}_2))(v_j(\vec{r}_1)-v_j(\vec{r}_2))\rangle\hat{z}_i\hat{z}_j\\
      =&2[(B(0)-B(r,\mu))+(C(0)-C(r,\mu))\cos^2\gamma\\
      &-A(r,\mu)\cos^2\theta-2D(r,\mu)\cos\theta\cos\gamma]
    \end{aligned}
\end{equation}
which is determined by the coefficients $A(r,\mu)$, $B(r,\mu)$, $C(r,\mu)$, $D(r,\mu)$, and the angle $\theta$ between the LOS and $\vec{r}$. We list the coefficients in Appendix.~\ref{appendix.A} and the derivation can be found in \citet{2016MNRAS.461.1227K}. By performing Legendre expansion for the coefficients up to the second order, i.e.,  $A(r,\mu)=\sum_nA_n(r)P_n(\mu)\approx A_0(r)+A_2(r)P_2(\mu)$, and using the relation $\mu(\gamma,\theta,\phi)=\sin\gamma\sin\theta\cos\phi+\cos\gamma\cos\theta$, $D_z(\vec{r})$ can be further simplified to:
\begin{equation}
\begin{aligned}
\label{eq.Dzr}
D_z(\vec{r})&\approx c_1-c_2\cos\phi-c_3\cos^2\phi\\
\end{aligned}
\end{equation}
The parameters $c_1$, $c_2$, and $c_3$, which absorb the dependence on $\gamma$ and $\rm M_A$, are listed in Appendix.~\ref{appendix.A} for Alfv\'en mode. As we can write $|\vec{r}|=\sqrt{R^2+z^2}$ and $\tan\theta=\frac{R}{z}$, the parameters only depends on $R$, $z$, $\gamma$, $\rm M_A$. By performing integration along z-direction, the resulting isotropy degree ${\rm iso(R,\gamma, M_A, \Delta v)}$ is only the function of $R$, $\gamma$, $\rm M_A$, and $\Delta v$. 

Here we normalize $\Delta v$ so that its maximum value is 1. In the limit cases of thin channel $\Delta v =0 $ and thick channel $\Delta v =1 $, defining:
\begin{equation}
\begin{aligned}
&\alpha(R,\phi)=[D_z(\vec{r})+2\beta_T]^{-1/2},\\
&\chi(R,\phi)=\alpha(0,0)-\alpha(R,\phi),\\
\end{aligned}
\end{equation}
the corresponding isotropy degree can be expressed as:
\begin{equation}
\begin{aligned}
&{\rm iso(R, \gamma, M_A,\Delta v=0)}\\
&=\frac{\int_{-S}^{S}\chi(R, 0)dz}{\int_{-S}^{S}\chi(R,\pi/2)dz};\\
&{\rm iso(R, \gamma, M_A,\Delta v=1)}\\
&=\frac{\int_{-S}^{S}\chi(R,0)dz\int_{-1/2}^{+1/2} W(v_{los})\exp[-\frac{1}{2}v_{los}^2\alpha^2(R,0)]dv_{los}}{\int_{-S}^{S}\chi(R,\pi/2)dz\int_{-1/2}^{+1/2} W(v_{los})\exp[-\frac{1}{2}v_{los}^2\alpha^2(R,\pi/2)]dv_{los}};\\
\end{aligned}
\end{equation}
\begin{figure}
\centering
\includegraphics[width=1.0\linewidth,height=0.687\linewidth]{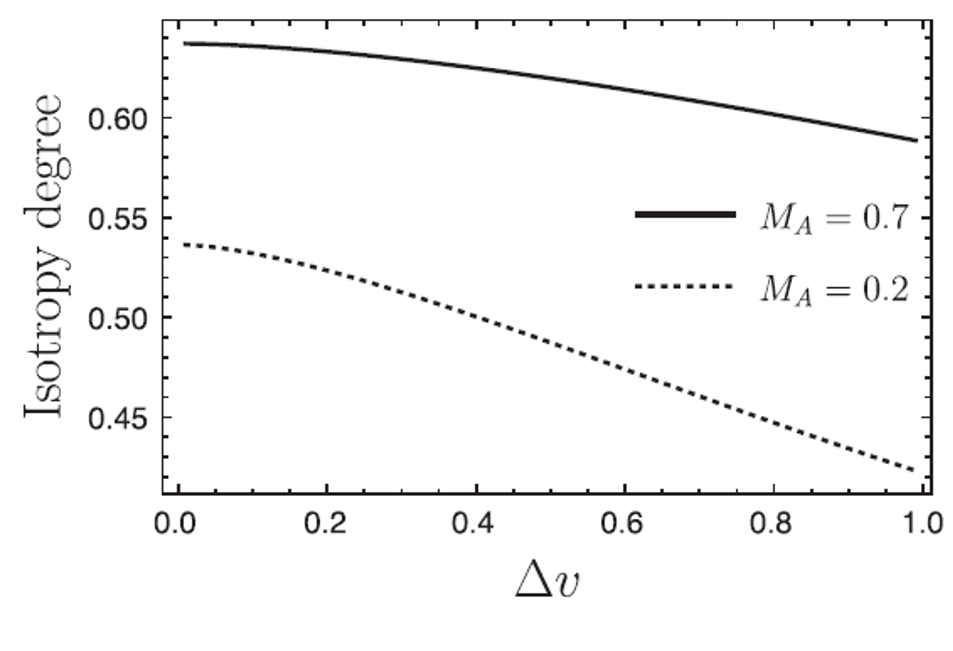}
\caption{\label{fig:kandel} An illustration of the variation of isotropy degree with respect to velocity channel width $\Delta v$ and $\rm M_A$ at $\gamma=\pi/2$ considering only Alfv\'enic wave. Extracted from \citet{2016MNRAS.461.1227K}.}
\end{figure}

Note that in the global reference frame, the anisotropy is scale-independent. It means the denominator and nominator of ${\rm iso(R,\gamma, M_A, \Delta v)}$ are both proportional to $\propto R^a$, where $a$ is a constant determined by turbulence's properties (as $D_z(\vec{r})\propto r^\nu$; see Appendix.~\ref{appendix.A} and \citealt{2016MNRAS.461.1227K}). Consequently, the measured ${\rm iso(\gamma, M_A, \Delta v=0)}$ and ${\rm iso(\gamma, M_A, \Delta v=1)}$ only depend on $\gamma$ and $\rm M_A$. The two values of ${\rm iso(\gamma, M_A, \Delta v)}$, therefore, are sufficient to determine $\gamma$ and $\rm M_A$ for a given system.
\begin{figure*}
\centering
\includegraphics[width=0.9\linewidth,height=0.74\linewidth]{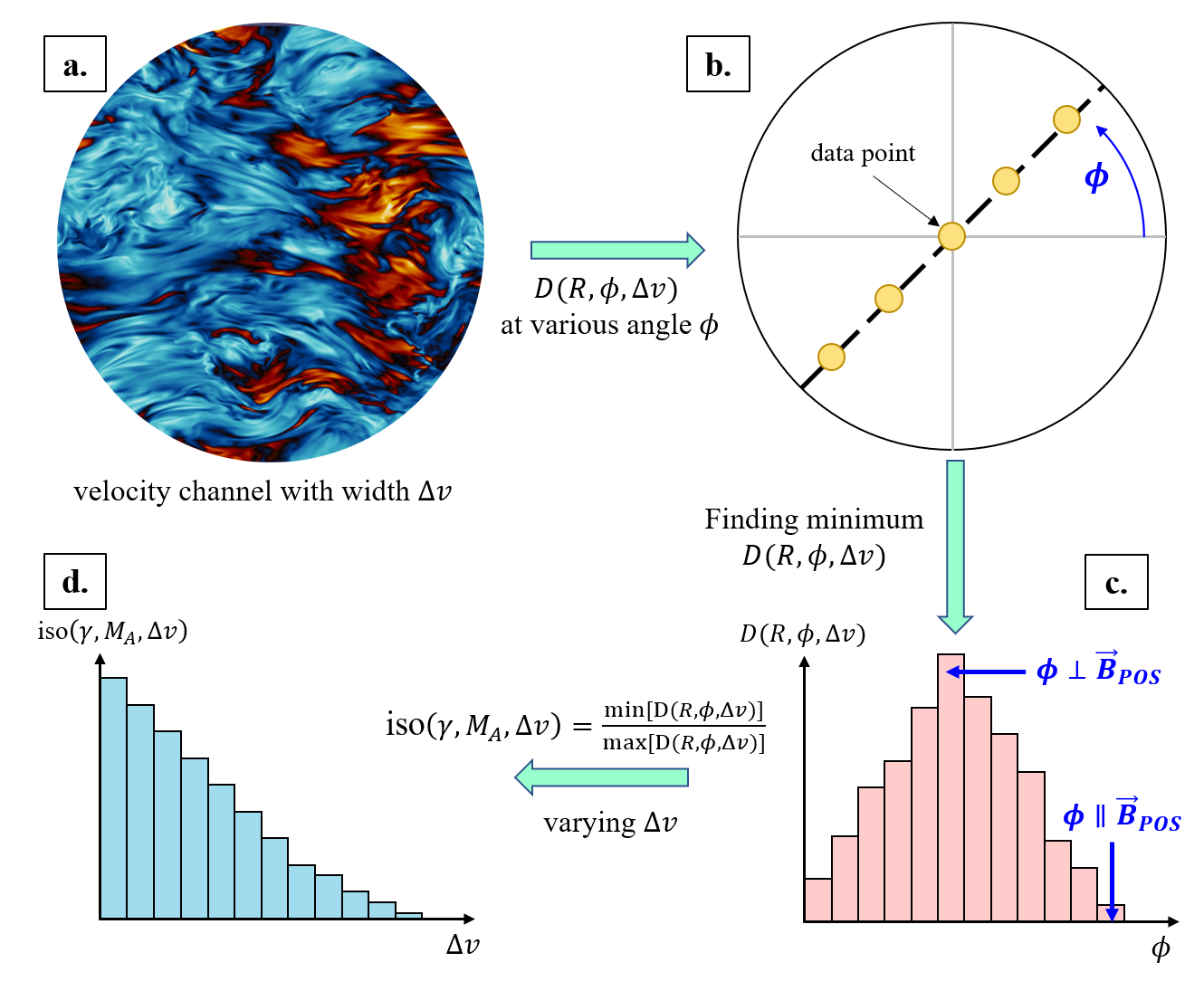}
\caption{\label{fig:recipe} Diagram of the SFA procedure to trace three-dimensional magnetic fields. \textbf{Step a:} choosing a velocity channel with width $\Delta v$. \textbf{Step b:} calculating the structure function $D(R,\phi,\Delta v)$ at a given distance $R$ and position angle $\phi$. Repeating this step by varing $\phi$ from 0 to $2\pi$. \textbf{Step c:} plotting the relation of $D(R,\phi,\Delta v)$ and $\phi$. The angle $\phi$ corresponding to {\it maximum} $D(R,\phi,\Delta v)$ gives the direction perpendicular to the POS magnetic field. The angle $\phi$ corresponding to {\it minimum} $D(R,\phi,\Delta v)$ gives the direction parallel to the POS magnetic field. \textbf{Step d:} finding the isotropy degree $\rm iso(\gamma,M_A,\Delta v)$ from step c. Plotting the relation of $\rm iso(\gamma,M_A,\Delta v)$ and $\Delta v$ by repeating steps a-c for various $\Delta v$. Solving the value of $\gamma$ and $\rm M_A$ from the maximum and minimum values of $\rm iso(\gamma,M_A,\Delta v)$.}
\end{figure*}

Considering that the intensity structure function $D(R,\phi,\Delta v)$ is proportional to $\cos^{-1}\gamma$, $\cos^{-2}\gamma$, and $\rm M_A^{-2/3}$ (as $D_z(\vec{r})$ depends on $\cos^{-2}\gamma$, $\cos^{-4}\gamma$, and $\rm M_A^{-4/3}$; see Appendix.~\ref{appendix.A}), we separate the variables into\footnote{we drop higher orders as $|\cos\gamma|\le1$ and $\rm M_A\le1$ for sub-Alfv\'en turbulence}:
\begin{equation}
\label{eq.iso}
    {\rm iso(\gamma, M_A, \Delta v})=(a_1+a_2\cos\gamma+a_3\cos^{2}\gamma)(b_1+b_2{\rm M_A^{2/3}})f(\Delta v)
\end{equation}
where $a_1$, $a_2$, $a_3$, $b_1$, $b_2$ are parameters to be determined and $f(\Delta v)$ is a function of $\Delta v$. Note that the dependence on $\gamma$, which only involve in the projection of structure function (see Eq.~\ref{eq.9}), is the same for all MHD modes, i.e., Alfv\'en, fast, and slow modes. The $\rm M_A$ term comes from the amplitude of the power spectrum (see Appendix.~\ref{appendix.A}). Nevertheless, the power spectrum of slow mode is the same as that of Alfv\'en mode  and fast mode is isotropic in the zeroth approximation (see \citealt{2016MNRAS.461.1227K} and Appendix.~\ref{Appendix:C}).
%{\al it is good to present the structure functions of the three modes in the global system of ref. for a couple of $M_A$. The fast modes, as we know, have the anisotropy perpendicular to the Alfven modes (see Lazarian \& Pogosyan 2012). It is important to determine what accuracy the "zeroth approximation" has.} 
Therefore, we expect the fitting function works for the mixture of all MHD modes. As an example shown in Fig.~\ref{fig:kandel}, at $\gamma=\pi/2$ and constant density, ${\rm iso(\gamma, M_A, \Delta v})$ is negatively related to $\Delta v$, but positively related to $\rm M_A$. In the following, we will perform numerical study to determine the parameters.

%1. $R=0$ means $\phi=\theta=0$:
%\begin{equation}
%\begin{aligned}
%\int_{-S}^S\alpha(0,0)dz=
%&\int_{-S}^S\frac{dz}{\sqrt{(q_1+q_2+q_3)r^\nu+2\beta_T}}\\
%=&\int_{-S}^S\frac{dz}{\sqrt{(q_1+q_2+q_3)z^\nu+2\beta_T}}
%\end{aligned}
%\end{equation}

%2. $R\ne0, \phi=0, \theta\ne0$:
%\begin{equation}
%\begin{aligned}
%\int_{-S}^S\alpha(R,\pi/2)dz&=\int_{-S}^S\frac{dz}{\sqrt{D_z(\vec{r}_{R,0})+2\beta_T}}\\
%D_z(\vec{r}_{R,0})/r^\nu&=(q_1+q_2\cos^2\theta+q_3\cos^4\theta)\\
%&\times[1-\frac{(s_1+s_2\cos^2\theta)\sin\theta\cos\theta\sin\gamma\cos\gamma}{q_1+q_2\cos^2\theta+q_3\cos^4\theta}\\
%&-\frac{(u_1+u_2\cos^2\theta)\sin^2\theta\sin^2\gamma}{q_1+q_2\cos^2\theta+q_3\cos^4\theta}]
%\end{aligned}
%\end{equation}
    
%3. $R\ne0,\phi=\pi/2,\theta\ne0$:
%\begin{equation}
%\begin{aligned}
%&\int_{-S}^S\alpha(R,0)dz=
%\int_{-S}^S\frac{dz}{\sqrt{(q_1+q_2\cos^2\theta+q_3\cos^4\theta)r^\nu+2\beta_T}}\\
%&=\int_{-S}^S\frac{dz}{\sqrt{(q_1+q_2\frac{z^2}{R^2+z^2}+q_3\frac{z^4}{(R^2+z^2)^2})(R^2+z^2)^{\nu/2}+2\beta_T}}
%\end{aligned}
%\end{equation}
\section{Methodology}
\label{sec.method}
\subsection{Structure function analysis}
The calculation of isotropy degree is performed by the second-order structure-function, which is called the Structure-Function Analysis (SFA; \citealt{SFA,XH21}).

The first step is to determine the POS magnetic field. As illustrated in Fig.~\ref{fig:recipe}, we choose a velocity channel which has an arbitrary width $\Delta v$. For this velocity channel, we calculate the intensity structure-function $D(R,\phi,\Delta v)$ :
\begin{equation}
        D(R,\phi,\Delta v)=\langle|I(\boldsymbol{X_1})-I(\boldsymbol{X_2})|^2 \rangle
\end{equation}
where $\boldsymbol{X_1}$ and $\boldsymbol{X_2}$ denote the 2D coordinates of two intensity data points locating at position angle $\phi$. Note this calculation is preformed in the global reference frame, which means the anisotropy is scale-independent. Consequently at arbitrary $R$, $ D(R,\phi,\Delta v)$ always exhibits maximum value when $\phi$ is perpendicular to the POS magnetic field direction and minimum value when $\phi$ is parallel to the POS magnetic field direction. Therefore, the POS magnetic field direction can be determined by varying the position angle from 0 to $\pi$. Note in numerical simulations, there exists a fake dependence of the anisotropy on length scales, due to the isotropic driving and insufficient inertial range \citep{2000ApJ...539..273C,2018ApJ...865...54Y}. To avoid this fake anisotropy, one should select $R$ at sufficiently small scales away from the driving scale. In our case, we have $R=10$ pixels (see also Fig.~\ref{fig:vda}).

The second step is to figure out the LOS magnetic field direction and 3D magnetization, i.e., the inclination angle $\gamma$ and 3D $\rm M_A$. Here we define the istropy degree $\rm iso(\gamma,M_A,\Delta v)=\max[D(R,\phi,\Delta v)]/\min[D(R,\phi,\Delta v)]$. To extract $\gamma$ and $\rm M_A$, one needs at least two measurements of $\rm iso(\gamma,M_A,\Delta v)$. For instance, one choose to measure $\rm iso$ at normalized channel width $\Delta v\approx1$ and $\Delta v\approx0$. By solving Eq.~\eqref{eq.iso}, one can get the values of $\gamma$ and $\rm M_A$.

\subsection{Velocity Decomposition Algorithm}
%{\al If you are not using thermal broadening, this should be explained, e.g. you are using heavy species that are being moved by hydrogen. In fact, the VDA is designed to deal with thermal broadening. Therefore without this effect taken into account, the improvement can be marginal.}
In Sec.~\ref{sec:theory}, we discussed the pure Alfv\'en mode case neglecting the compressible components, i.e., fast and slow modes, as well density fluctuations. This simplification holds for subsonic turbulence, in which the density field passively regulating by velocity field follows the statics of velocity fluctuation \citep{2005ApJ...624L..93B,2019ApJ...878..157X}. However, for supersonic turbulence, the turbulent compression is more significant, and the presence of shocks modifies the density field's statics \citep{2007ApJ...658..423K,H2,XH21}. The contribution from density can further change the anisotropy of the observed intensity structure. It is, therefore, essential to remove density's contribution from channel maps.

In addition to the density effect, in a real scenario, one has to consider the thermal broadening effect in the velocity channel map, particularly for warm media. In this work, the temperature in simulations is set to 10 K, so the thermal broadening is marginal. %which corresponds to heavy molecules moved by hydrogen.
Nevertheless, \citet{2020arXiv201215776Y} recently developed a novel technique, i.e., the Velocity Decomposition Algorithm (VDA), to reduce the density and thermal broadening effect in a given channel map. We briefly discuss the recipe of VDA here, and more details can be found in \citet{2020arXiv201215776Y}. 

For a given intensity field $\rho(x,y,v_{los})$ in PPV space (see Eq.~\eqref{eq.max}), we can define the integrated intensity map $I(x,y)$ (i.e., the moment-0 map) and velocity channel map $Ch(x,y)$ as:
\begin{equation}
\begin{aligned}
    I(x,y)&=\int_{-\infty}^{+\infty}\rho(x,y,v_{los})dv_{los}\\
    Ch(x,y)&=\int_{v_0-\Delta v/2}^{v_0+\Delta v/2}\rho(x,y,v_{los})dv_{los}
\end{aligned}
\end{equation}
where $v_0$ is the velocity of the averaged emission line maximum. Velocity fluctuation dominates the observed intensity fluctuations in the channel map when channel width $\Delta v$ satisfies $\Delta v<\sqrt{\delta( v_{los}^2)}$, where $\sqrt{\delta(v_{los}^2)}$ is the velocity dispersion \citep{LP00}. Accordingly, the velocity contribution $Ch_v(x,y)$ and density contribution $Ch_d(x,y)$ in $Ch(x,y)$ can be extracted from  \citet{2020arXiv201215776Y}:
\begin{equation}
\begin{aligned}
    Ch_v(x,y)&=Ch-(\langle Ch\cdot I\rangle-\langle Ch\rangle\langle I\rangle)\frac{I-\langle I\rangle}{\sigma_I^2}\\
    Ch_d(x,y)&=(\langle Ch\cdot I\rangle-\langle Ch\rangle\langle I\rangle)\frac{I-\langle I\rangle}{\sigma_I^2}\\
\end{aligned}
\end{equation}
where $\sigma_I^2=\langle(I-\langle I\rangle)^2\rangle$ and $\langle...\rangle$ denotes the ensemble average. 

\section{Numerical Data}
\label{sec:data}
\begin{table}
\centering
\label{tab:sim}
\begin{tabular}{| c | c | c | c | c |}
\hline
Model & $\rm M_S$ & $\rm M_A$  & Resolution & $\beta$\\\hline \hline
A1 & 0.66 & 0.12 & $792^3$ & 0.07 \\
A2 & 0.63 & 0.34 & $792^3$ & 0.58 \\
A3 & 0.62 & 0.56 & $792^3$ & 1.63 \\
A4 & 0.60 & 0.78 & $792^3$ & 3.38 \\
A5 & 0.60 & 1.02 & $792^3$ & 5.78 \\
A6 & 0.89 & 0.54 & $792^3$ & 0.74 \\\hline
B1 & 0 & 0.8 & $512^3$ & $\infty$ \\
B2 & 0 & 3.2 & $512^3$ & $\infty$ \\\hline
C1 & 10.81 & 0.26 & $792^3$ & 0.001 \\
C2 & 11.12 & 0.37 & $792^3$ & 0.002 \\
C3 & 10.53 & 0.51 & $792^3$ & 0.005 \\
\hline
\end{tabular}
\caption{Description of our MHD simulations. $\rm M_S$ and $\rm M_A$ are the instantaneous values at each the snapshots are taken. The compressibility of turbulence is characterized by $\beta=2(\frac{\rm M_A}{\rm M_S})^2$.}
\end{table}
\begin{figure*}
\centering
\includegraphics[width=1.0\linewidth,height=0.63\linewidth]{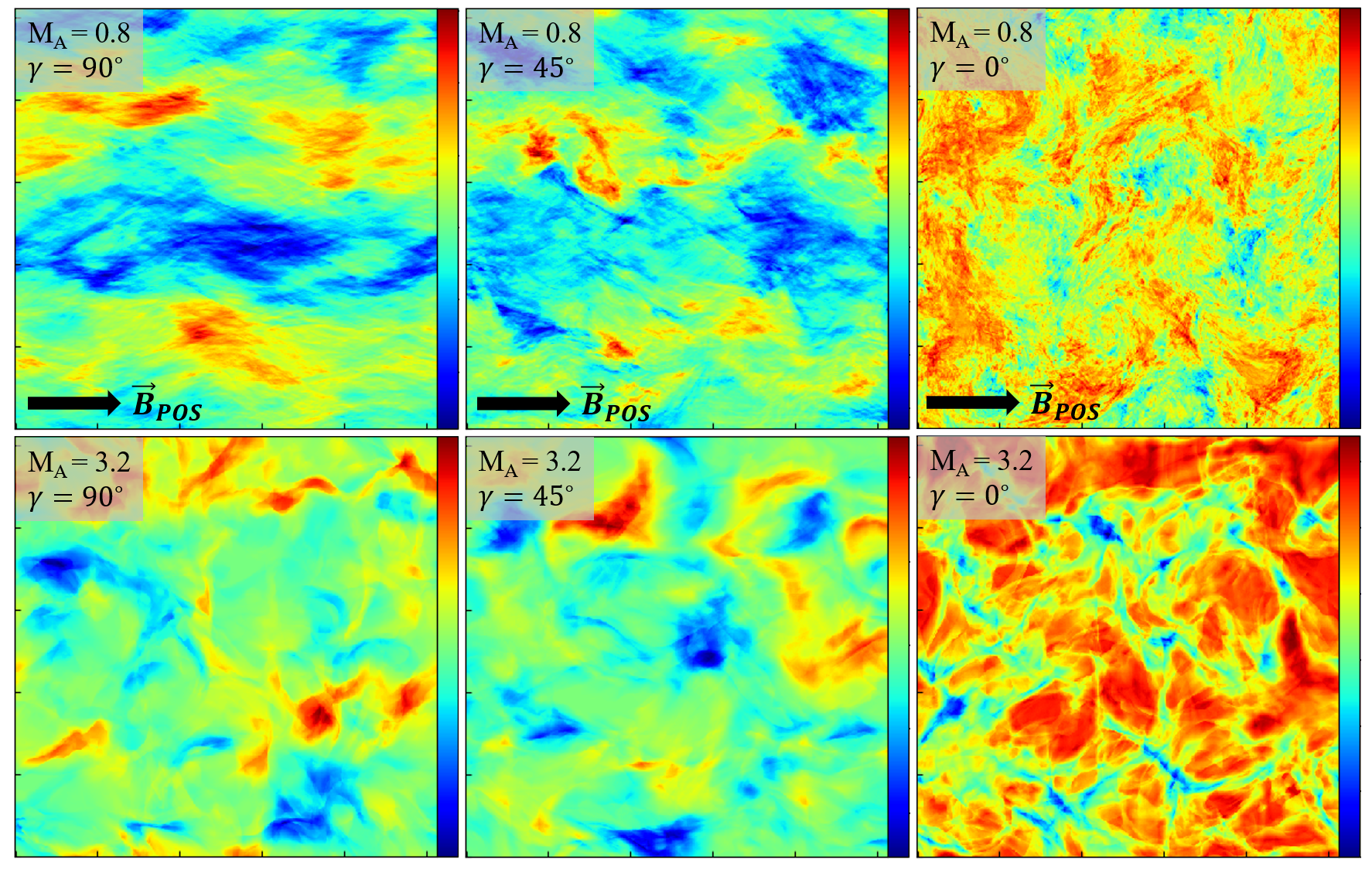}
\caption{\label{fig:map}The velocity channel maps of incompressible simulations $\rm M_A=0.8$ (top) and $\rm M_A=3.2$ (bottom). We use normalized velocity channel width $\Delta v=0.1$.}
\end{figure*}

\subsection{Incompressible MHD simulations}
The 3D incompressible MHD simulations are produced from a pseudospectral code developed by \citet{2000ApJ...539..273C}. The code is a third-order-accurate hybrid employing Essentially Non-Oscillatory (ENO) scheme. It uses hyperviscosity and hyperdiffusivity to solves the periodic incompressible MHD equations in a periodic box of size $2\pi$:
\begin{equation}
	\begin{aligned}
		\frac{\partial \Vec{v}}{\partial t}&=(\nabla \times \Vec{v})\times\Vec{v}-v_A^2(\nabla\times\Vec{B})\times\Vec{B}+\nu\nabla^2\Vec{v}+\Vec{f}+\nabla P'\\
		\frac{\partial \Vec{B}}{\partial t}&=\Vec{B}\cdot\nabla\Vec{v}-\Vec{v}\cdot\nabla\Vec{B}+\eta\nabla^2\Vec{B}
	\end{aligned}
\end{equation}
where $\Vec{f}$ is the random driving force and $P'=P+ \Vec{v}\cdot\Vec{v}/2$ is the pressure. $\nu$ and $\eta$ represent kinematic viscosity and magnetic diffusivity ($\eta=\nu=6.42\times10^{-32}$) . The magnetic field is considered as $\Vec{B}=\Vec{B}_0+\delta\Vec{B}$, where $\Vec{B}_0$ is the uniform background field and $\delta\Vec{B}$ is fluctuation. $\Vec{B}_0$ initially is perpendicular to the LOS. $v_A$ is the Alfv\'en speed of the uniform background. The code employs a pseudospectral method. We calculate the MHD equations in real space and transform them into Fourier space to obtain the Fourier components of nonlinear terms. The calculation of the temporal evolution is performed in Fourier space. Turbulence is solenoidally driven by 21 forcing components with $2<k<\sqrt{12}$ resulting in a peak of energy injection at $k \approx 2.5$. Each forcing component has a correlation time of one. In our case, we vary the value of $\Vec{B}_0$ to achieve $\rm M_A=0.8$ and $\rm M_A=3.2$ and stagger the simulation to 512$^3$ cells/pixels. The respective parameters are listed in Tab.~\ref{tab:sim}. We refer the reader to \citet{2000ApJ...539..273C} and \citet{2010ApJ...725.1786C} for further details.

\subsection{Compressible MHD simulation}
We generate 3D compressible MHD simulations through ZEUS-MP/3D code \citep{2006ApJS..165..188H}, which solves the ideal MHD equations in a periodic box. \begin{equation}
	\begin{aligned}
		&\partial\rho/\partial t +\nabla\cdot(\rho\Vec{v})=0\\
		&\partial(\rho\Vec{v})/\partial t+\nabla\cdot[\rho\Vec{v}\Vec{v}+(P+\frac{B^2}{8\pi})\Vec{I}-\frac{\Vec{B}\Vec{B}}{4\pi}]=\Vec{f}\\
		&\partial\Vec{B}/\partial t-\nabla\times(\Vec{v}\times\Vec{B})=0
	\end{aligned}
\end{equation}
where $\Vec{f}$ is a random large-scale driving force, $\rho$ is the density, $\Vec{v}$ is the velocity, and $\Vec{B}$ is the magnetic field. We also consider a zero-divergence condition $\nabla \cdot\Vec{B}=0$, and an isothermal equation of state $P = c_s^2\rho_0$, where $P$ is the gas pressure and $c_s\approx0.192$ (in numerical unit) is the sound speed. The magnetic field and density field are considered as $\Vec{B}=\Vec{B}_0+\delta \Vec{b}$ and $\rho=\rho_0+\delta\rho$, where $\Vec{B}_0$ and $\rho_0$ are the uniform background fields. $\delta\rho$ and $\delta\Vec{b}$ represent fluctuations. Initially, $\Vec{B}_0$ is assumed to be perpendicular to the LOS.

We consider single fluid and operator-split MHD conditions in the Eulerian frame. The simulation is staggered to 792$^3$ cells/pixels, and turbulence is solenoidally injected at wavenumber k $\approx$ 2 in Fourier space. The turbulence gets numerically dissipated at wavenumber k $\approx$ 100.

As a scale-free MHD turbulence simulation is only characterized by the sonic Mach number M$_{\rm S}=v_{inj}/c_{s}$ and Alfv\'{e}nic Mach number M$_{\rm A}=v_{inj}/v_{A}$, were $v_{inj}$ is the isotropic injection velocity, and $v_{A}$ is the Alfv\'{e}n speed. We vary the value of $\Vec{B}_0$ and $\rho_0$ to achieve different $\rm M_{\rm S}$ and $\rm M_{\rm A}$. The parameters are listed in Tab.~\ref{tab:sim}. In the text, we refer to the simulations by either the model name or the physical parameters.

%%%%%%%%%%%%%%%%%%%%%%%%%%%%%%%%%%%%%%%%%%%%%%%%%%%%%%%%%%%%%%%%%%%%%%%%%%%%%%%%%%
\section{Results}
\label{sec:results}
\subsection{Incompressible MHD turbulence}
\label{subsec:POS}
\begin{figure*}
\centering
\includegraphics[width=0.8\linewidth,height=0.7\linewidth]{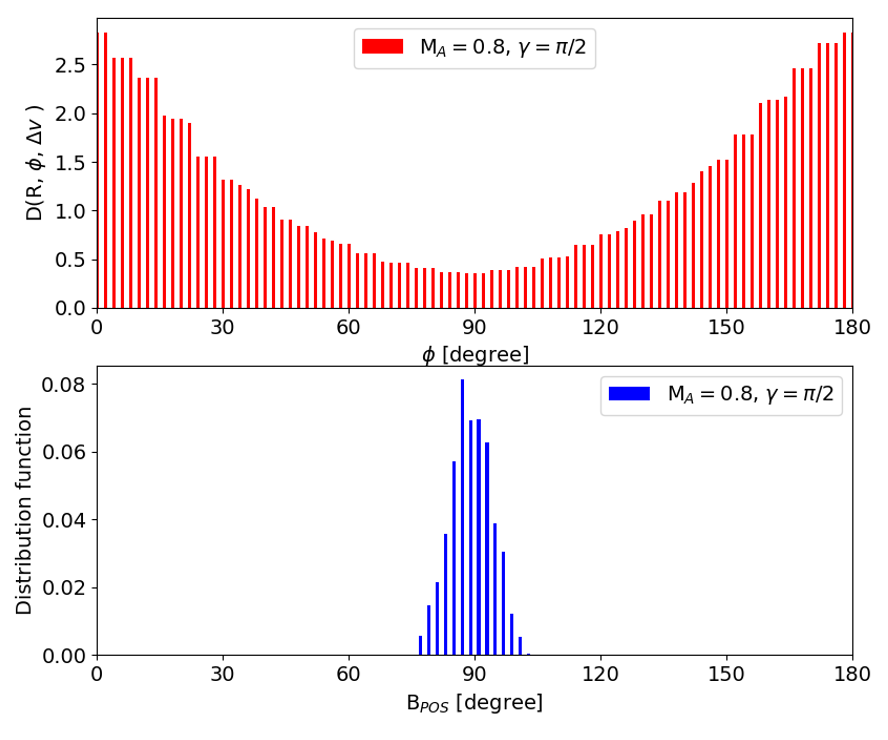}
\caption{\label{fig:DvB}\textbf{Top:} The correlation of the structure function $\rm D(R,\phi,\Delta v)$ and position angle $\phi$. We use incompressible simulation $\rm M_A=0.8$ and choose $\Delta v=0.1$, $R=10$ pixels, and $\gamma=\pi/2$. \textbf{Bottom:} The histogram of the POS magnetic field direction in IAU convention.}
\end{figure*}

\begin{figure}
\centering
\includegraphics[width=1.0\linewidth,height=0.73\linewidth]{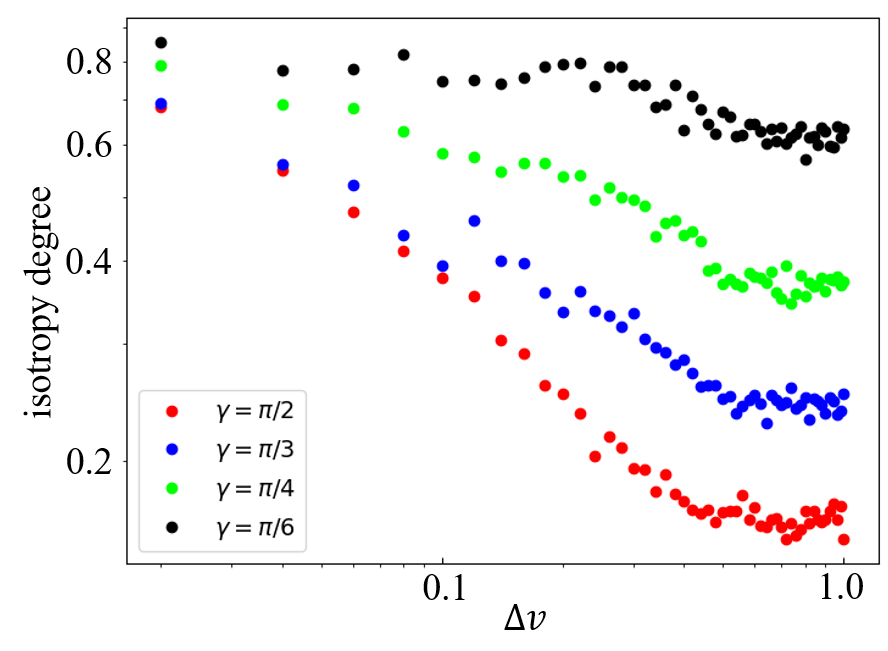}
\caption{\label{fig:incompress}The correlation of isotropy degree with respect to normalized velocity channel width $\Delta v$. The incompressible simulation $\rm M_A=0.8$ is used here.}
\end{figure}

We firstly examine the application of SFA to incompressible turbulence. Fig.~\ref{fig:map} presents the velocity channel maps (normalized $\Delta v=0.1$) of incompressible simulations $\rm M_A=0.8$ and $\rm M_A=3.2$ at various inclination angles of mean magnetic fields. When the total mean magnetic field is perpendicular to the LOS (i.e., $\gamma=\pi/2$), the channel map's intensity structures ($\rm M_A=0.8$) are dominated by striations aligned with the POS magnetic field. The decreasing inclination angle, however, diminishes the anisotropic striation. When the total mean magnetic field is parallel to the LOS, the intensity structures become isotropic. As for sup-Alfv\'enic turbulence, the intensity structures are always isotropic. In Fig.~\ref{fig:DvB}, we calculate the intensity structure-function $D(R,\phi,\Delta v)$ using the incompressible simulation $\rm M_A=0.8$ and choosing $\Delta v=0.1$, $R=10$ pixels, and $\gamma=\pi/2$. We vary the position angle $\phi$ from 0 to 180$^\circ$. The maximum value of $D(R,\phi,\Delta v)$ appears at 0, and 180$^\circ$, while the minimum value appears at 90$^\circ$. Note there is a 180$^\circ$ ambiguity. Also, from the histogram of the POS magnetic field direction, we find the magnetic field direction concentrates at 90$^\circ$ with a mean value $\approx89.96^\circ$. $\phi=90^\circ$ which corresponds to the minimum $D(R,\phi,\Delta v)$, therefore, gives the direction of the mean POS magnetic field. 

\begin{figure}
\centering
\includegraphics[width=1.0\linewidth,height=0.76\linewidth]{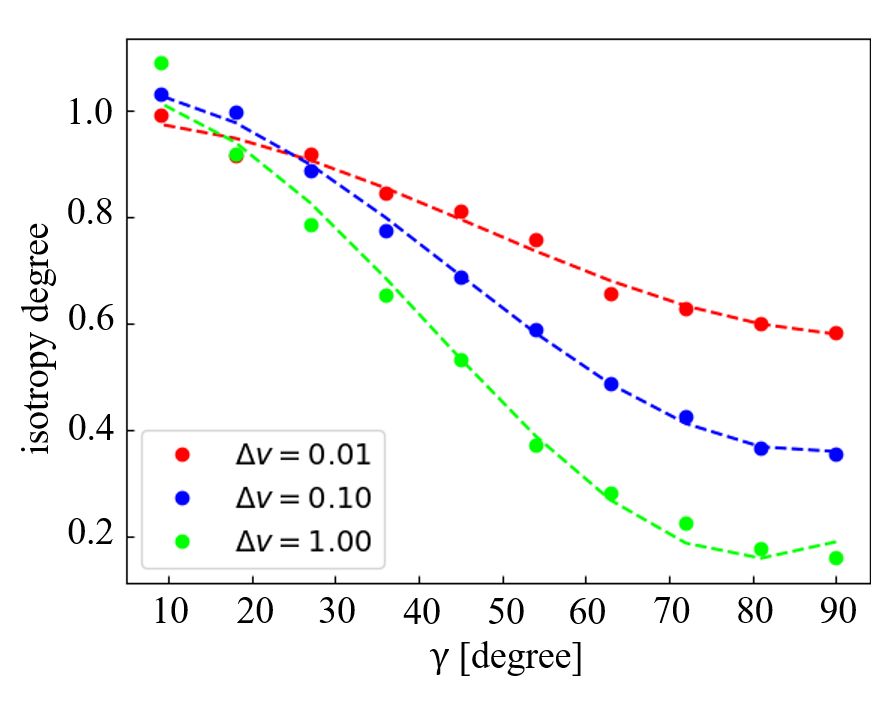}
\caption{\label{fig:JCho_v}The correlation of isotropy degree with respect to inclination angle $\gamma$. The incompressible simulation $\rm M_A=0.8$ is used here. The dashed lines represent the fitting model $\rm iso(\gamma,M_A,\Delta v)=a'_1+a'_2\cos\gamma+a'_3\cos^{2}\gamma$.}
\end{figure}

Furthermore, we rotate the simulation cube so that the relative angle between the mean magnetic field and the LOS is $\gamma$. By varying the normalized channel width $\Delta v$, we plot the isotropy degree corresponding to different inclination angles in Fig.~\ref{fig:incompress}. Its uncertainty is given by the standard error of the mean, which is negligible here due to large sample size of the entire cube. We find the isotropy degree generally decreases when the channel becomes thick. The maximum and minimum values appear at normalized $\Delta v\approx0.01$ and $\Delta v\approx1$, respectively. This decrease can be understood as all thin channel emitters have similar LOS velocities, and anisotropy is suppressed. Also, when $\gamma$ is smaller, i.e., the mean magnetic field is more parallel to the LOS, the observed anisotropy gets smaller as well. This decrease with respect to $\gamma$ comes from the fact that the anisotropy is less projected onto the POS.

In Fig.~\ref{fig:JCho_v}, we fix the normalized channel width $\Delta v$ to be 0.01, 0.10, and 1.00 but varying the value of $\gamma$, which is uniformly spaced in [0, 90$^\circ$]. We can see the isotropy degree is decreases when $\gamma$ becomes larger, since more anisotropy is projected onto the POS. In the case of $\gamma<10^\circ$, the isotropy degree is 1, which indicates that the velocity channel is isotropic. In addition, we find the data points well fit the model $\rm iso(\gamma,M_A,\Delta v)=a'_1+a'_2\cos\gamma+a'_3\cos^{2}\gamma$. Note here we already fixed $\Delta v$ and $\rm M_A=0.8$. The fitting parameters are:
\begin{enumerate}
    \item  $\Delta v=0.01$: $a'_1=0.58\pm0.04$, $a'_2=0.07\pm0.18$, and $a'_3= 0.33\pm0.17$; 
    \item $\Delta v=0.10$: $a'_1=0.36\pm0.03$, $a'_2=-0.06\pm0.13$, and $a'_3= 0.75\pm0.12$;
    \item $\Delta v=1.00$:  $a'_1=0.19\pm0.08$, $a'_2=-0.39\pm0.38$, and $a'_3= 1.24\pm0.35$;
\end{enumerate}
We find $a'_1$ is small and $a'_3$ becomes large for a thick channel. It implies that the thick channel is more anisotropic and the $\gamma$ has more important role in regulating thick channel's intensity structure. 

%The reason on why there is a lower bound for $\gamma$ is because the turbulent component shall dominate over the mean field component when $\gamma<4\tan^{-1}(\rm M_A/3)$, resulting an underestimation of mean magnetic field strength in this regime
\subsection{Compressible MHD turbulence}

\begin{figure}
\centering
\includegraphics[width=1.0\linewidth,height=0.69\linewidth]{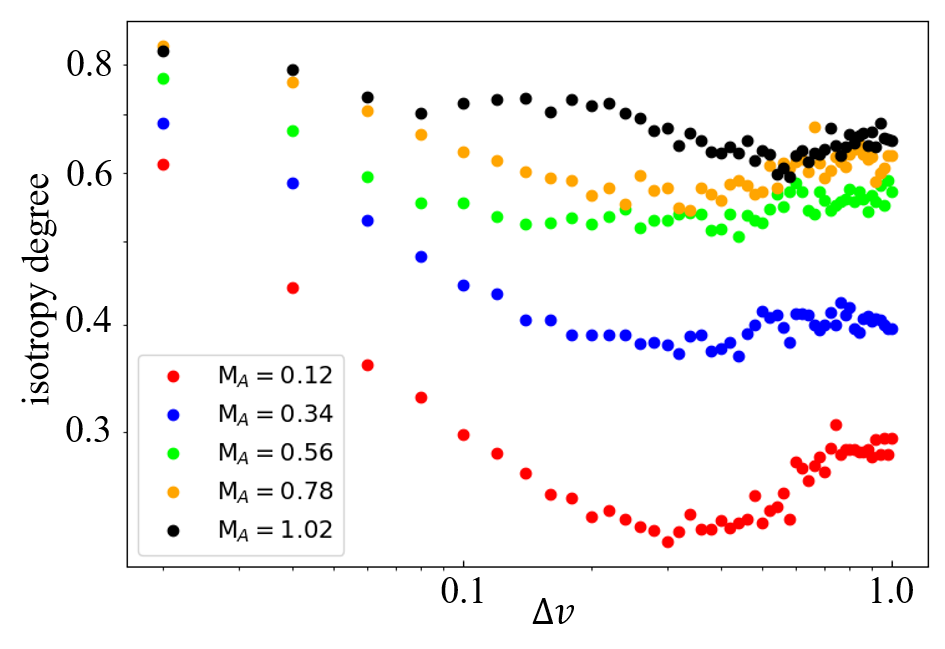}
\caption{\label{fig:sub_Ma}The correlation of isotropy degree with respect to normalized velocity channel width $\Delta v$ and $\gamma=\pi/2$. Compressible simulations $\rm M_S\approx0.6$ are used here.%{\al why does the red curve increases at large $\Delta v$?}
}
\end{figure}

In this section, we test the SFA using compressible MHD simulations. Unlike the incompressible case, compressible slow and fast modes as well density field start to affect the anisotropy. 

In Fig.~\ref{fig:sub_Ma}, we calculate the correlation of isotropy degree and normalized velocity channel width $\Delta v$ at $\gamma=\pi/2$. We find for sub-Alfv\'en turbulence, the isotropy degree is negatively related to $\Delta v\le0.3$. When $\Delta v\ge0.3$, the isotropy degree starts increasing, which indicates smaller anisotropy. This can be understood as the density fluctuation dominates the thick channel so that the anisotropy is diluted (see also Fig.~\ref{fig:iso_vda} and \citealt{LP00}). 
\begin{figure}
\centering
\includegraphics[width=1.0\linewidth,height=0.76\linewidth]{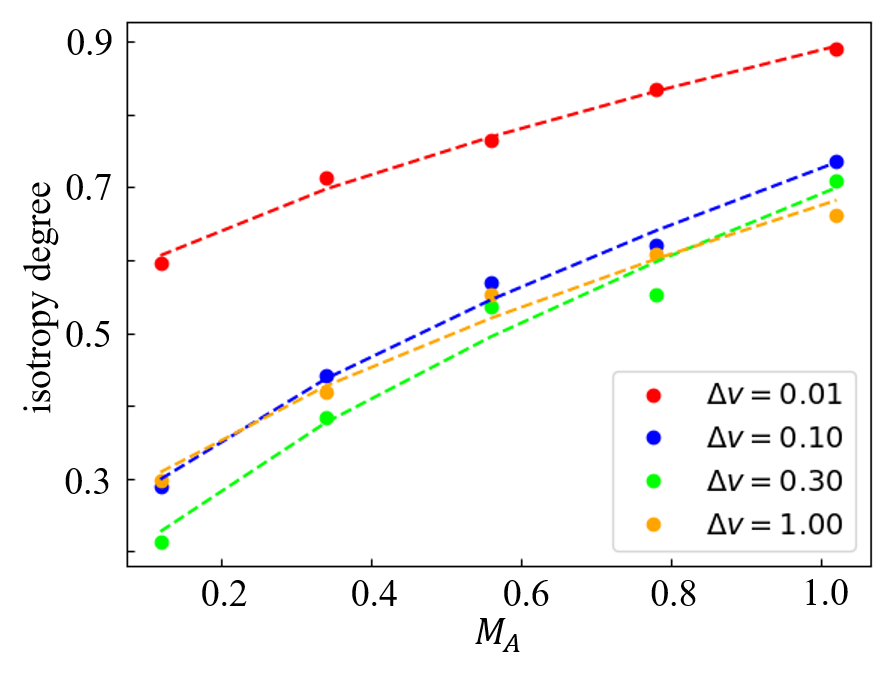}
\caption{\label{fig:sub_dv}The correlation of isotropy degree with respect to $\rm M_A$. The compressible simulations $\rm M_S\approx0.6$ and $\gamma=\pi/2$ are used here. The dashed lines represent the fitting model $\rm iso(\gamma,M_A,\Delta v)=b'_1+b'_2M_A^{2/3}$. %{\al why do the 3 lines cross. If the coefficients $b_1$ and $b_2$ are kept constant, this should not happen for the fitting formulae. What is wrong?}
}
\end{figure}

\begin{figure*}
\centering
\includegraphics[width=1.0\linewidth,height=1.11\linewidth]{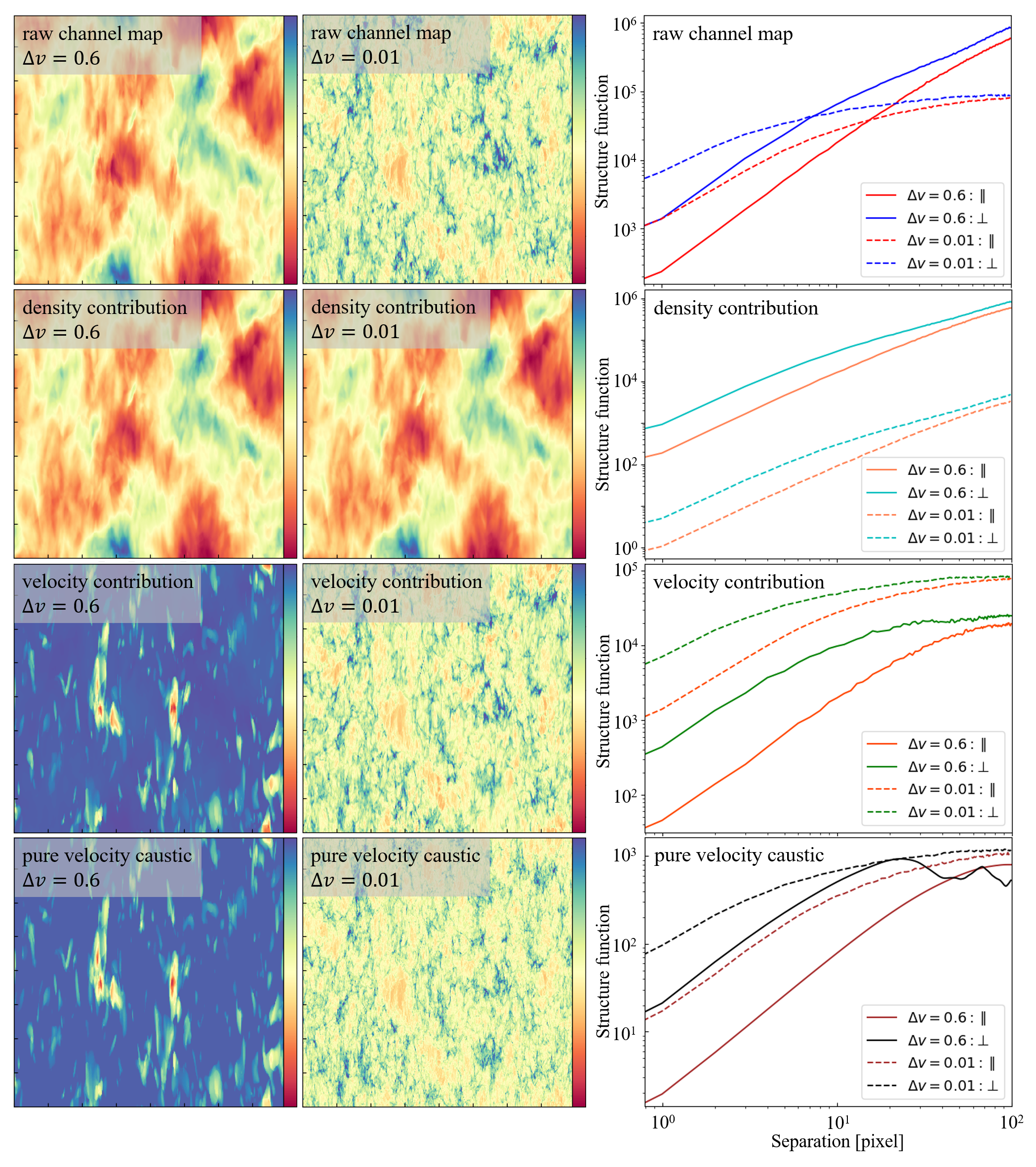}
\caption{\label{fig:vda} \textbf{The $1^{st}$ row:} The raw velocity channel maps (left and middle) and intensity structure functions (right) of compressible simulation $\rm M_S=0.66$ and $\rm M_A=0.12$. \textbf{The $2^{nd}$ row:} The density contribution extracted by VDA in the raw velocity channel maps and its intensity structure functions. \textbf{The $3^{rd}$ row:} The velocity contribution extracted by VDA in the raw velocity channel maps and its corresponding intensity structure functions. \textbf{The $4^{th}$ row:} The pure velocity caustic maps, i.e., setting a uniform density field when generating the PPV cube, and their corresponding intensity structure functions. Mean magnetic field is along the vertical direction. %{\al VDA in its present form is not expected to work for thick channels. Those get dominated by density. This should be discussed.}
}
\end{figure*}

In additional, in Fig.~\ref{fig:sub_dv}, we further fix the normalized channel width $\Delta v$ to be 0.01, 0.10, 0.30, and 1.00. We find it is clear that strong magnetic field cases exhibit more significant anisotropy. Sup-Alfv\'enic case is closer to be isotropic, as the intrinsic turbulence is isotropic. We also fit the data points with the model $\rm iso(\gamma,M_A,\Delta v)=b'_1+b'_2M_A^{2/3}$. Note here we already fixed $\Delta v$ and $\gamma=\pi/2$. The fitting parameters are:
\begin{enumerate}
    \item  $\Delta v=0.01$: $b'_1=0.52\pm0.04$ and $a'_2=0.37\pm0.06$; 
    \item $\Delta v=0.10$: $b'_1=0.16\pm0.07$ and $b'_2=0.57\pm0.10$;
    \item $\Delta v=0.30$: $b'_1=0.08\pm0.14$ and $b'_2=0.61\pm0.19$;
    \item $\Delta v=1.00$: $b'_1=0.19\pm0.09$ and $b'_2=0.49\pm0.13$;
\end{enumerate}
The models' good fitness (see also Fig.~\ref{fig:JCho_v}) with the data points confirms our theoretical expectation. We find that the fitted curve of $\Delta v=1.00$ gets crossed with other curves. We expect that this comes from density effect, as the $\Delta v=1.00$ case keeps only density fluctuations. We will study the effect of density contribution in the following section.

\subsubsection{Removing density contribution}
\label{subsub:vda}
\begin{figure}
\centering
\includegraphics[width=1.0\linewidth,height=0.75\linewidth]{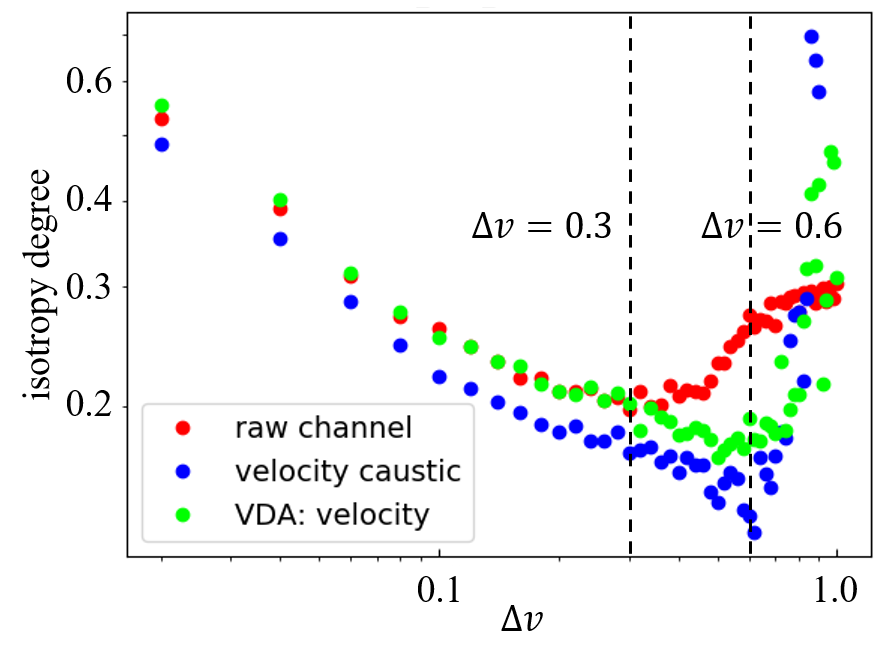}
\caption{\label{fig:iso_vda}The correlation of isotropy degree and normalized velocity channel width $\Delta v$. We make comparison for raw velocity channel (i.e., using real density field), pure velocity caustic (i.e., uniform density field), and the velocity contribution extracted by VDA. The compressible simulation $\rm M_S\approx0.6$, $\rm M_A\approx0.12$, and $\gamma=\pi/2$ is used.}
\end{figure}

From Eq.~\eqref{eq.cf}, it is clear that when the channel is thick, i.e., $\Delta v\to \infty$, the observed intensity fluctuations only include density's contribution \citep{LP04}:
\begin{equation}
\begin{aligned}
      \xi_I(R,\phi)\propto&\int_{-S}^{S}dz\frac{1+\widetilde{\xi}_\rho(\vec{r})}{\sqrt{D_z(\vec{r})+2\beta_T}}\int_{-\infty}^{+\infty}dv_{los} \exp[\frac{-v_{los}^2}{2(D_z(\vec{r})+2\beta_T)}]\\
      \propto&\int_{-S}^{S}dz[1+\widetilde{\xi}_\rho(\vec{r})]
\end{aligned}
\end{equation}
in which all the velocity information is erased, and density contribution plays a primary role in the observed intensity statistics. As velocity information is the most crucial in calculating the isotropy degree, we have to remove the density contribution in channel maps. To do so, here we use the VDA method (see \S~\ref{sec.method}). 
\begin{figure}
\centering
\includegraphics[width=1.0\linewidth,height=2.2\linewidth]{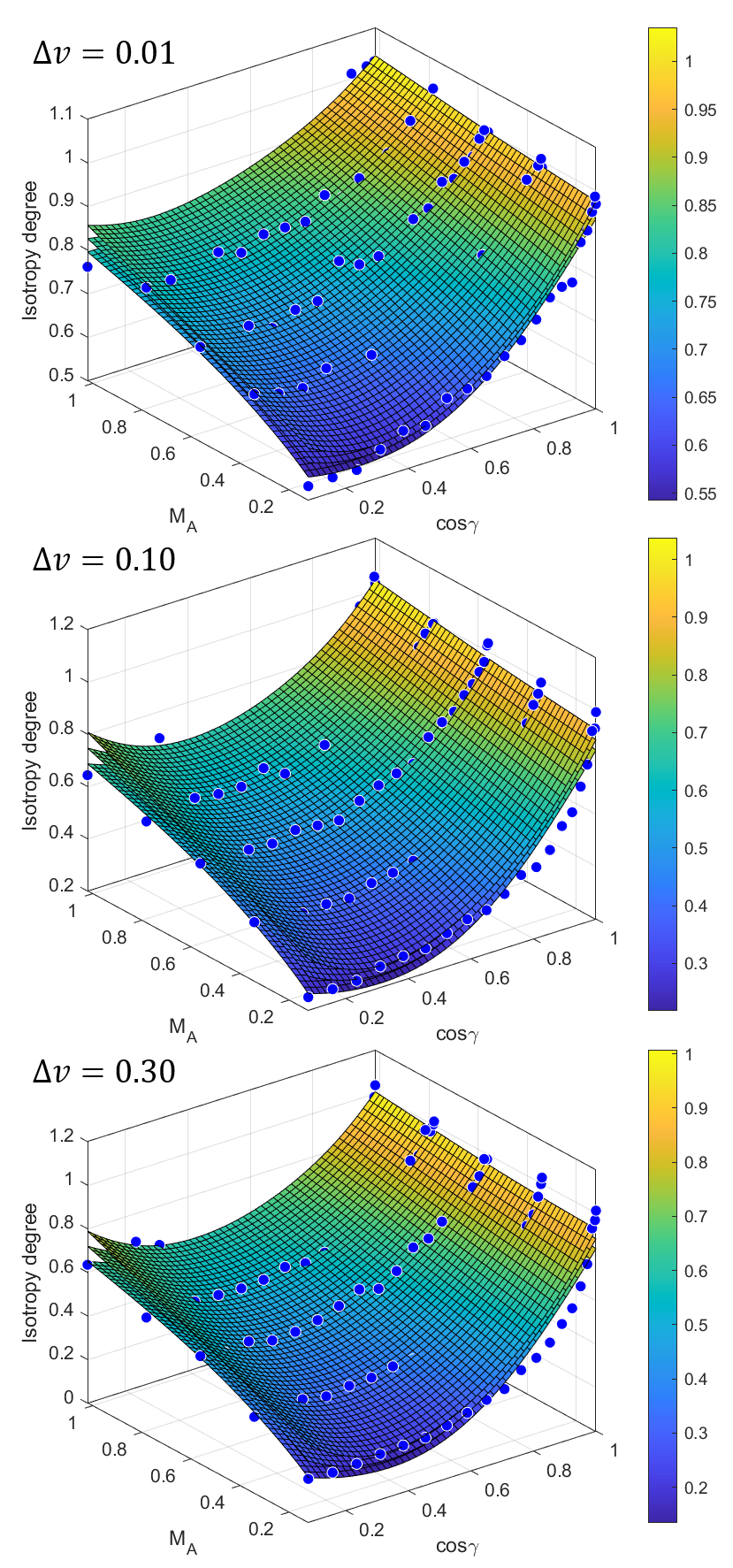}
\caption{\label{fig:sub_fitting}The correlation of isotropy degree, $\rm M_A$, and $\cos\gamma$. Dotted symbol denotes the raw data points (blue), which are fitted with a  model $\rm iso(\gamma,M_A,\Delta v)=a+b\cos\gamma+c\cos^2\gamma+d\cos\gamma M_A^{2/3}+eM_A^{2/3}+f\cos^2\gamma M_A^{2/3}$. Upper and lower layers give the fitting uncertainty. Compressible simulations $\rm M_S\approx0.6$ are used.}
\end{figure}

In Fig.~\ref{fig:vda}, we use the compressible simulation $\rm M_S=0.66$ and $\rm M_A=0.12$ choosing $\Delta v=0.6$ and $\Delta v=0.01$ at $\gamma=\pi/2$. Firstly, we plot the raw channel maps, and we find the thin channel map ($\Delta v=0.01$) appears more filamentary structures that are elongating along the mean magnetic field direction. We also calculate the intensity structure-function with respect to the mean magnetic fields' parallel and perpendicular directions. The structure functions get shallower for the thin channel map as more small-scale structures appear. After that, we decompose the velocity and density contributions from the raw channel maps with the VDA method. We find the for both thick and thin channel maps, their corresponding density contribution maps are highly similar. Importantly, the (raw) thick channel map's intensity structures show more similarity with the density contribution map. In contrast, the (raw) thin channel map is similar to the velocity contribution map. This is exactly the theoretical prediction of \citet{LP00}, i.e., the intensity fluctuations in thick and thin channels are dominated by density fluctuations and velocity fluctuations, respectively. 

Also, the velocity contribution map exhibit more significant anisotropy in terms of the structure functions, i.e., the larger difference between the parallel and perpendicular components. The density contribution map's structure functions show higher amplitude for the thick channel, as the density fluctuation is more significant in the thick channel. In contrast, the velocity contribution map's structure functions for the thin channel, in which the velocity fluctuation is more important. We also analyzed the pure velocity caustic effect by setting a uniform density field when generating a PPV cube \footnote{The concept of velocity caustics is firstly proposed by \citet{LP00} to signify the effect of density structure distortion due to turbulent velocities along the LOS. Setting a uniform density field, the intensity distribution in velocity channel maps is purely generated by velocity fluctuations.}. The VDA decomposed velocity contribution maps are highly similar to the pure velocity caustic maps, which confirm the validity of the VDA method. However, for the thick velocity caustic map, its structure-function starts oscillating when the separation is larger than 20 pixels. We find this comes from the fact that velocity information is marginal in a thick channel. For instance, for the pure velocity caustic case, because we set a uniform density field (i.e., $\rho(x,y,z)=1$) for a $792^3$ MHD simulation, the intensity value at the density dominated position would be saturated to 792, which erases all velocity information. These saturated intensity values are excluded when calculating the structure-function so that only velocity information is taken into account. The remaining velocity information does not guarantee an accurate structure-function at a large scale from what we see. This insufficiency of velocity information is more significant if the channel width increases further.

In Fig.~\ref{fig:iso_vda}, we make a comparison for the isotropy degrees calculated from raw channel map, velocity caustic map, and VDA decomposed velocity contribution map of simulation $\rm M_S=0.66$ and $\rm M_A=0.12$. We find three crucial $\Delta v$ ranges, at which the isotropy degree shows distinguishable behavior. For all three kinds of channel maps (i.e., raw channel map, velocity caustic map, and VDA decomposed velocity contribution map), the isotropy degree monotonically decreases when $\Delta v$ gets larger until $\Delta v\approx0.3$. However, for the raw channel map, the isotropy degree starts increasing when $0.3\le\Delta v\le0.6$. In contrast, the isotropy degrees of velocity caustic map and VDA velocity map keep decreasing until $\Delta v\approx0.6$. As discussed above, the velocity caustic map and VDA velocity map contain only velocity information while the raw channel map includes both density and velocity contribution. The difference of the isotropy degrees in the range of $0.3\le\Delta v\le0.6$ therefore comes from density contribution, which diminishes anisotropy. When $0.6\le\Delta v$, the isotropy degree of the raw channel map gets saturated since the projected density field's anisotropy is not related to channel width. The other two isotropy degrees, however, are dramatically diverged. This can be understood as the velocity information in a very thick channel is not statistically sufficient for structure function's calculation, i.e. the sample size is not enough. For instance, in Fig.~\ref{fig:vda}, the structure function of pure velocity caustic case  gets fluctuating when separation is larger than 20 pixels. When the channel becomes thicker, the remaining valid information of velocity fluctuations may drop down so that the isotropy degree at the 10-pixel scale (i.e., the numerical dissipation scale) gets diverged. Nevertheless, in observations, the inertial range and sample size are sufficiently large. One should perform the SFA at larger separations for thick channels to find sufficient velocity information.
%{\al the right procedure is to do structure function analysis for larger separations as the channels become thicker. The issue of thick or thin depends on the eddy size. We must go to larger eddies to keep thin.}
For supersonic turbulence, the density contribution is more significant. We discuss the case of supersonic turbulence in Appendix.~\ref{Appendix.B}.
\begin{table*}
\centering
\label{tab:param}
\begin{tabular}{| c | c | c | c | c | c | c |}
\hline
 $\Delta v$ & $a$ & $b$ & $c$ & $d$ & $e$ & $f$ \\\hline\hline
$\Delta v=0.01$ & 0.54 (0.51, 0.58)& -0.46 (-0.53, -0.39) & 0.86 (0.80, 0.91) & 0.26 (0.23, 0.29) & 0.29 (0.28, 0.31) & -0.47 (-0.49, -0.45)\\
$\Delta v=0.10$ & 0.26 (0.19, 0.34) & -0.89 (-1.03, -0.75) & 1.47 (1.37, 1.58) & -0.03 (-0.06, 0.06) & 0.54 (0.51, 0.57) & -0.39 (-0.43, -0.43)\\
$\Delta v=0.30$ & 0.22 (0.14, 0.39) & -1.12 (-1.23, -0.97) & 1.76 (1.65, 1.88) & 0.05 (-0.02, 0.12) & 0.57 (0.53, 0.60) & -0.51 (-0.56, -0.46)\\
$\Delta v=0.60$ & 0.20 (0.09, 0.32) & -1.08 (-1.29, -0.87) & 1.68 (1.52, 1.84) & -0.04 (-0.09, 0.09) & 0.58 (0.53, 0.63) & -0.49 (-0.56, -0.42)\\
\hline
\end{tabular}
\caption{The coefficients for fitting model $\rm iso(\gamma,M_A,\Delta v)=a+b\cos\gamma+c\cos^2\gamma+dM_A^{2/3}\cos\gamma +eM_A^{2/3}+fM_A^{2/3}\cos^2\gamma $. The upper and lower bounds within 95\% confidential level of the fitting model are provided in brackets. The compressible simulations $\rm M_S\approx0.6$ are used for the fitting.}
\end{table*}
\subsection{Determining $\gamma$ and $\rm M_A$}
The POS magnetic field direction can be traced by varying the position angle used for calculating the intensity structure-function (see \S~\ref{subsec:POS}). Combining the model of $\rm iso(\gamma,M_A,\Delta v)$ determined in this section, one can further access the inclination angle $\rm \gamma$ and the total Alfv\'en Mach number $\rm M_A$. The three-dimension magnetic field information, including both orientation and strength, is achievable in PPV space. 

As the isotropy degree $\rm iso(\gamma,M_A,\Delta v)$ for a given $\Delta v$ depends only on $\gamma$ and $\rm M_A$, two measurements of $\rm iso(\gamma,M_A,\Delta v)$ are sufficient to determine an unique pair of $\gamma$ and $\rm M_A$, although multiple measurements could reduce uncertainty. Fig.~\ref{fig:sub_fitting} considers a fitting model $\rm iso(\gamma,M_A,\Delta v)=a+b\cos\gamma+c\cos^2\gamma+d M_A^{2/3}\cos\gamma +eM_A^{2/3}+f M_A^{2/3}\cos^2\gamma$, which comes the expansion of Eq.~\ref{eq.iso}. We selected four normalized channel widths $\Delta v=0.01, 0.10, 0.30, 0.60$. We vary the values of $\gamma$ for each compressible simulation $\rm M_S\approx0.6$. By performing a two-variable fitting, we find out the coefficients and list them in Tab.~\ref{tab:param}. We find the terms $\cos\gamma$, $\cos^2\gamma$, and $\rm M_A^{2/3}$ have higher weights in a thick channel. It means the anisotropy in a thick channel is more sensitive to $\cos\gamma$. Also, the weights of $\rm M_A^{2/3} \cos\gamma$ term are close to zeros when $\Delta v>0.01$, which means their contribution is negligible.  Note that density has an important role in thick channel $\Delta v\ge0.3$. When calculating the isotropy degree, it is advantageous to take thin channel width at the range of $\Delta v\le0.3$.

We test our fitting model in the compressible simulation $\rm M_S\approx0.89$, $\rm M_A\approx0.54$. We rotate the simulation so that the mean magnetic field is inclined to the LOS with angle $\gamma$. Following the recipe illustrated in Fig.~\ref{fig:recipe}, we firstly determine the mean POS magnetic field direction. We use the \textbf{Alignment Measure} (AM, \citealt{2017ApJ...835...41G}): 
\begin{align}
\label{AM_measure}
AM=2\cos^{2} \theta_{r}-1
\end{align}
to quantify the relative angle $\theta_r$, which is the difference between the orientation of the magnetic field inferred from the SFA and the real magnetic field. AM's value is in the range of [-1, 1]. AM = 1 means the difference between the two vector's orientations is zero which AM = -1 indicates a 90$^\circ$ difference. To measure the mean POS magnetic field, we vary the position angle $\phi$ from 0 to $180^\circ$ with resolution $1^\circ$. The position angle corresponding to the minimum value of the intensity structure-function reveals the POS magnetic field direction. Also, here we select the velocity channel with widths $\Delta v=0.01$ and $\Delta v=0.10$. 
\begin{figure}
\centering
\includegraphics[width=1.0\linewidth,height=1.34\linewidth]{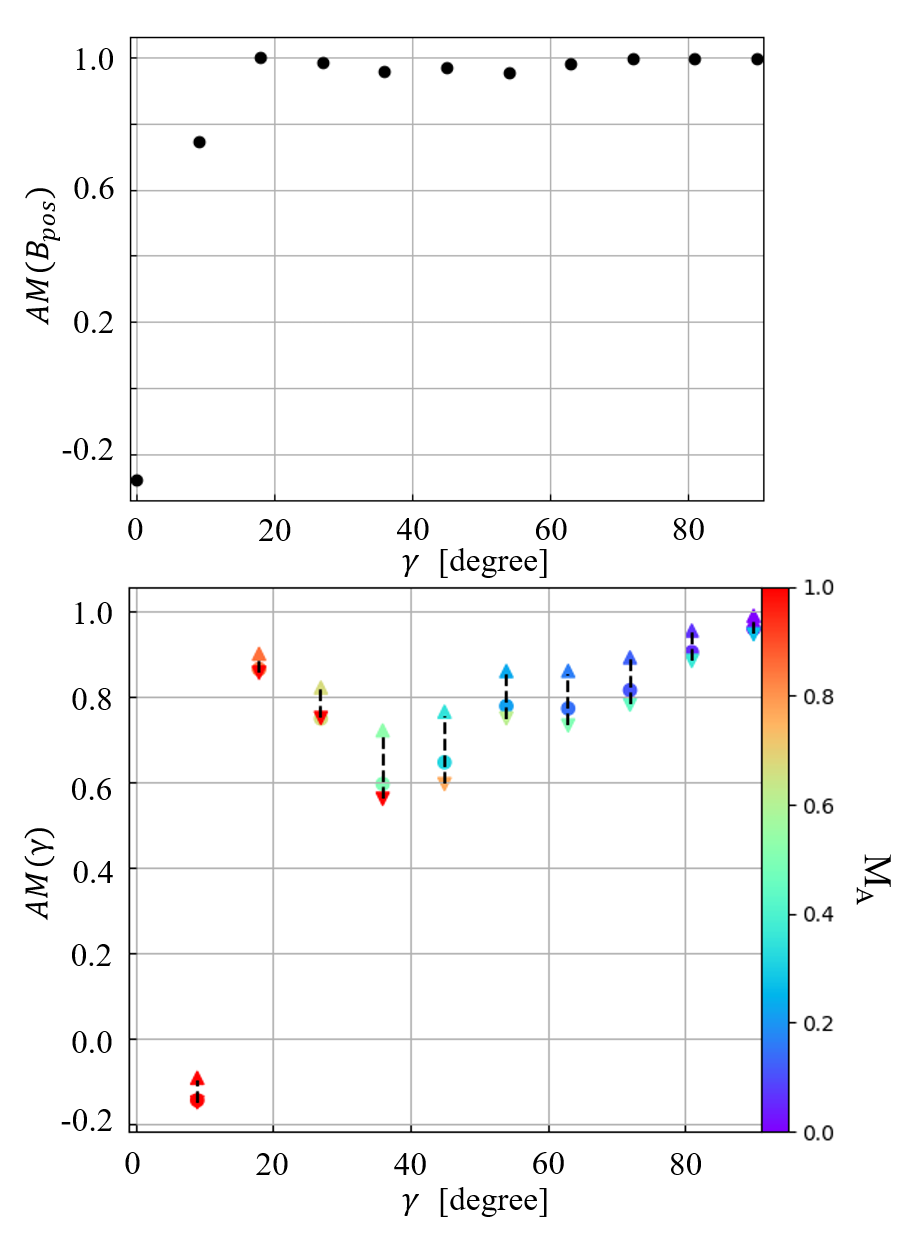}
\caption{\label{fig:sub08}\textbf{Top:} The AM of measured $B_{pos}$ and real $B_{pos}$ in cases of different inclination angle $\gamma$ (x-axis). \textbf{Bottom:} The AM of measured $\gamma$ and real $\gamma$ (x-axis). Color of the point represents the measured $\rm M_A$ for the simulation $\rm M_S\approx0.89$, $\rm M_A\approx0.54$. Circular symbol is the measured value and triangular symbol gives the uncertainty coming from the parameters used in fitting model.}
\end{figure}

In Fig.~\ref{fig:sub08}, we find the measured POS magnetic fields (using $\Delta v=0.01$) have excellent agreement (AM $\approx1$) with the real POS magnetic fields when $\gamma$ is larger than $18^\circ$. when $\gamma$ is close to zeros, the AM becomes negative. This can be understood from the fact that the POS magnetic field is isotropic in $\gamma\approx0$. Consequently, the intensity structure-function gets similar results along all directions, which cannot determine the POS magnetic field. With the knowledge of the POS magnetic field direction derived from the SFA, one can further calculate the isotropy degree, which is defined as the ratio of the intensity structure functions measured in the direction parallel and perpendicular to the POS magnetic field. We calculate the isotropy degrees at $\Delta v=0.01$ and $\Delta v=0.10$ and then solve the fitting model $\rm iso(\gamma,M_A,\Delta v)=a+b\cos\gamma+c\cos^2\gamma+d M_A^{2/3}\cos\gamma +eM_A^{2/3}+f M_A^{2/3} \cos^2\gamma $ %using the "vpasolve" function provided by MATLAB. 
The uncertainties are given by the upper and lower limits of the fitting model listed in Tab.~\ref{tab:param}. As shown in Fig.~\ref{fig:sub08}, the SFA measured $\gamma$ gives $\rm AM\ge0.6$ when the real $\gamma$ is larger than $18^\circ$. In the case of $\gamma<18^\circ$, the AM dramatically drops down to negative and $\rm M_A$ is underestimated. The misalignment and underestimation come from the fact that the turbulent components dominate over the mean-field components at small $\gamma$. This effect is more significant in estimating total magnetic field strength. The bound for this underestimation theoretically is $\gamma<4\tan^{-1}({\rm M_A}/\sqrt{3})$ (see \citealt{2020arXiv200207996L}). Also, we find the AM gets smaller than 0.8 when $\gamma$ is in the range of [36$^\circ$, 63$^\circ$]. This comes from the deviation of the measured POS magnetic field, which affects the resulting isotropy degree. As the measured $\gamma$ and $\rm M_A$ highly depend on the isotropy degree, a small fluctuation in the isotropy degree can lead to a significant deviation. 

\subsection{Effect of noise}
\begin{figure}
\centering
\includegraphics[width=1.0\linewidth,height=0.75\linewidth]{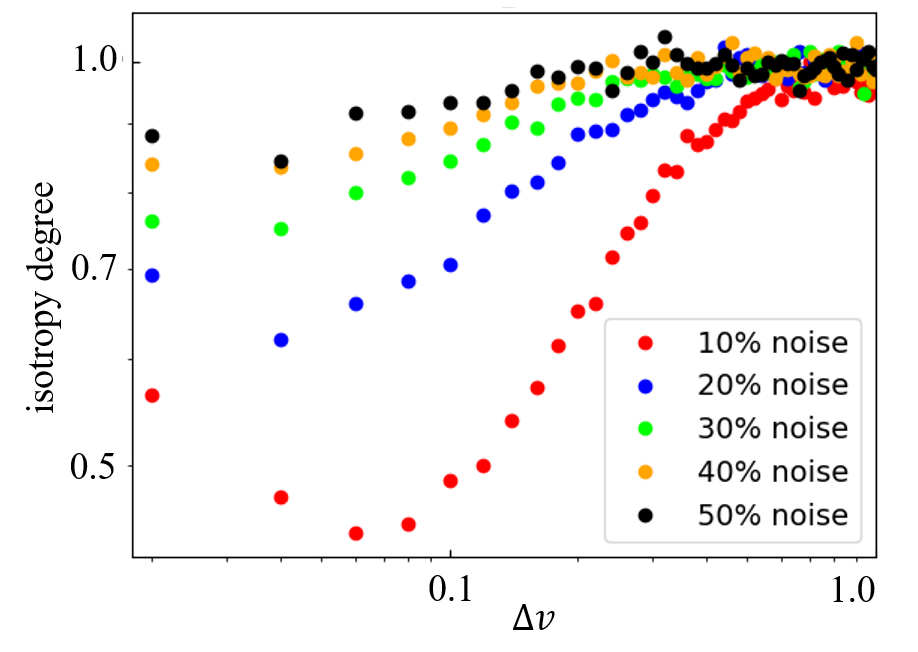}
\caption{\label{fig:noise}The correlation of isotropy degree and normalized channel width $\Delta v$ in the presence of noise. The noise's amplitude vary from 10\% to 50\% of the channel's mean intensity value. The compressible simulation $\rm M_S\approx0.6$, $\rm M_A\approx0.12$, and $\gamma=\pi/2$ is used.}
\end{figure}
Here we investigate the noise effect (i.e., the signal-to-noise ratio) on the measured isotropy degree. We use the compressible simulation $\rm M_S\approx0.6$, $\rm M_A\approx0.12$ as an example. We add Gaussian noise to each channel map. The noise amplitude varies from 10\% to 50\% of the channel's mean intensity value. In Fig.~\ref{fig:noise}, we find the isotropy degree has two distinguishable behaviors. Starting from the thinnest case, the isotropy degree decreases. Then when $\Delta v$ surpasses a turning point, the isotropy degree becomes positively related to $\Delta v$. The turning point, however, get small for a strong noisy case, since more velocity fluctuation is overwhelmed by noise. Also, in all cases, the isotropy degree converges to value 1.0 at $\Delta v=1.0$, which means the intensity map is isotropic. Nevertheless, even in the presence of noise, one should always choose a channel as thin as possible to calculate the isotropy degree.

%%%%%%%%%%%%%%%%%%%%%%%%%%%%%%%%%%%%%%%%%%%%%%%%%%%%%%%%%%%%%%%%%%%%%%%%%%%%%%%%%%
\section{Discussion}
\label{sec.dis}
\subsection{Comparison with other works}
Probing magnetic fields in ISM is notoriously challenging. One of the possibilities to measure the three-dimensional magnetic fields is using polarized dust thermal emission \citet{2019MNRAS.485.3499C}. This method measures the inclination angle $\gamma$ from:
\begin{equation}
    \sin^2\gamma=\frac{p_{obs}(1+\frac{2}{3}p_0)}{p_0(1+p_{obs})}
\end{equation}
where $p_{obs}$ is the observed polarization fraction and $p_0=\frac{3p_{max}}{3+p_{max}}$, in which $p_{max}$ is the maximum polarization fraction for a given cloud. It gives an integrated measurement of the magnetic field along the LOS.  

To access the local three-dimensional magnetic fields, the anisotropic properties of MHD turbulence provides one solution. For instance, the fluctuation of synchrotron polarization has the imprint of magnetic fields. Using multiple-wavelengths measurements of synchrotron polarization and taking the gradients of its wavelength derivative, the POS and LOS magnetic fields' orientation can be retrieved accordingly \citep{LY18b}. This method is called the Synchrotron Polarization Derivative Gradients (SPDGs). 

Another possibility of tracing three-dimension magnetic fields through the anisotropy is enabled by young stars' 3D velocity field, which is available in the GAIA survey \citep{Ha20}. As the turbulent velocities measured in the directions perpendicular and parallel to the magnetic field can be significantly different, the value of velocity's structure-function could reveal the three-dimensional magnetic field \citep{SFA}. Notably, the ratio of turbulent velocities measured in perpendicular and parallel directions of the magnetic field is proportional to $\rm M_A^{-4/3}$. The three-dimensional magnetic field's orientation and strength can therefore be determined simultaneously. This method is called the Structure-Function Analysis (SFA)

This work further extends the applicability of SFA to the spectroscopic PPV cubes. Here we measure the intensity structure-function in velocity channel maps with various channel widths. We find the isotropy degree, which is the ratio of intensity structure-functions measured in parallel and perpendicular directions of the POS magnetic fields, satisfies $\rm iso(\gamma,M_A,\Delta v)=a+b\cos\gamma+c\cos^2\gamma+d\cos\gamma M_A^{2/3}+eM_A^{2/3}+f\cos^2\gamma M_A^{2/3}$. The coefficients are listed in Tab.~\ref{tab:param}. By measuring the intensity structure function and isotropy degrees at two channel widths, one can simultaneously obtain the POS magnetic field, the magnetic field's inclination angle, and the total magnetization. Comparing with the dust polarization method and the SPDGs, the SFA gives complete measurements of three-dimensional magnetic fields, including both orientation and strength. 

\subsection{Prospects for the SFA}
The SFA method in revolutionary changes the stage of magnetic field measurements in ISM. It provides the possibility to simultaneously reveal the POS magnetic field, the LOS magnetic field, and total magnetic field strength using only a single spectroscopic PPV cube. This three-dimensional measurement of magnetic field benefits several studies.

The SFA is beneficial in modeling the foreground magnetic field and predicting the foreground dust polarization, which is indispensable in detecting the B-mode polarization of the inflationary gravitational wave. Previously, several efforts, including the Rolling Hough Transform (RHT; see \citealt{2015PhRvL.115x1302C} and \citealt{2019ApJ...887..136C}) and the Velocity Gradients Technique (VGT; see \citealt{EB} and \citealt{2020MNRAS.496.2868L}), have been made to predict the foreground dust polarization through atomic \ion{H}{1} gas. These predictions, however, consider only the POS magnetic fields neglecting that inclination of the magnetic field, which is one of the significant sources of depolarization. The simplification, thus, results in a higher value of the predicted polarization fraction. However, our method can provide a considerably accurate foreground dust polarization map by incorporating the LOS magnetic fields and improving on these existing methods.

On the other side, the relative role of turbulence, magnetic fields, self-gravity in star formation is a subject of intensive debate. Earlier measurements of the magnetic field are limited to two-dimensional. The POS magnetic fields in molecular clouds are usually inferred from either far-infrared polarimetry \citep{Aetal15,2016A&A...596A.105P} or VGT \citep{survey,velac,IRAM,2020arXiv200715344A}. The Zeeman splitting gives the signed LOS magnetic field strength \citep{2012ARAA...50..29}. Comparing to these methods, the SFA directly gives three-dimensional magnetic fields, which is crucial for an advanced understanding of star formation. The SFA can also be employed to study the magnetic fields in the galaxy's disk, where traditional far-infrared polarimetry suffers from distinguishing multiple clouds along the same LOS.

\subsection{Scope of pure MHD turbulence simulation}
In ISM, turbulence is ubiquitous, and its injection scale $L_{inj}$ is quite large. For example, \cite{2004ARA&A..42..211E} and \citet{2010ApJ...710..853C} reported $L_{inj}\approx100$ pc. Also, for diffuse ISM, the self-gravity is negligible so that turbulence and magnetic field are dominated there. These conditions well match with the scale-free simulations of pure MHD turbulence used in this work. Therefore, our analysis is appropriate for diffuse ISM regime.

At small scales, additional effects might become important. For instance, in star-forming regions, strong self-gravity, interstellar feedback, or outflows can be dominated. The stellar winds, jets, or bubbles can also change the picture of MHD turbulence. Consequently, additional studies are required for investigating turbulence at the scales or regions where these effects are important. 

\subsection{Implication for gradient studies}
The Velocity Gradients Technique (VGT; see \citealt{2017ApJ...835...41G, YL17a, LY18a,PCA}) is a new technique developed to trace the magnetic fields. As its theoretical foundation is the anisotropy of MHD turbulence, several properties of SFA can also immigrate to VGT. For example, the velocity gradient's direction indicates the eddy's semi-minor axis, and the velocity gradient's dispersion is related to eddy's size, i.e., the anisotropy. Since the eddy's size is regulated by $\rm M_A$ (see Fig.~\ref{fig:Ma}), velocity gradient's dispersion can reveal the magnetization \citep{2018ApJ...865...46L}. As we showed in this work, the anisotropy in PPV space is a function of channel width $\Delta v$, $\rm M_A$, and inclination angle $\gamma$. The velocity dispersion is therefore correlated to these three variables. By analogy to SFA, we expect two measurements of velocity gradient's dispersion at two given channel widths can uniquely determine a pair of $\rm M_A$ and $\gamma$. 

On the other hand, $\gamma$ is crucial for achieving 3D magnetic field strength. \citet{2020arXiv200207996L} recently propose a new technique to evaluate the POS magnetic field from two Mach numbers, i.e.,the sonic one $\rm M_S$ and the Alfv\'en one $\rm M_A$:
\begin{equation}
B=\Omega c_s\sqrt{4\pi\rho_0}{\rm M_S M_A^{-1}},
\end{equation}
where $\Omega$ is a geometrical factor ($\Omega=1$ corresponds magnetic fields perpendicular to the LOS), $c_s$ is sound speed, and $\rho_0$ is mass density.
In the frame of VGT, the measurements of $\rm M_S$ and $\rm M_A$ come from velocity gradients' amplitude \citep{YL20a} and dispersion \citep{2018ApJ...865...46L}, respectively. Consequently, with the knowledge of $\gamma$, VGT can also achieve a measurement of three-dimensional magnetic fields, including direction and strength. 

%%%%%%%%%%%%%%%%%%%%%%%%%%%%%%%%%%%%%%%%%%%%%%%%%%%%%%%%%%%%%%%%%%%%%%%%%%%%%%%%%%
\section{Summary} 
\label{sec:con}
The three-dimensional magnetic field is not directly achievable in observation. In this work, based on MHD turbulence's anisotropy, we develop a new technique (i.e., the SFA) to extract the three-dimensional magnetic field in PPV space. To sum up:
\begin{enumerate}
    \item We confirm that the anisotropy of observed intensity structures in PPV space is regulated by channel width $\Delta v$, inclination angle $\gamma$ of the magnetic field relative to the LOS, and Alfv\'en Mach number $\rm M_A$.
    \item We find the isotropy degree of intensity structure-function is anti-correlated with the width of velocity channels.
    \item We develop an algorithm in tracing three-dimensional magnetic field:
    \begin{enumerate}
        \item The intensity structure-function measures the POS magnetic field in a thin channel. The position angle, which minimizes the intensity structure-function, reveals the POS magnetic field direction;
        \item The inclination angle $\gamma$ and total Alfv\'en Mach number $\rm M_A$ are determined by the isotropy degrees of intensity structure-functions at two channel widths.
    \end{enumerate}
    \item We construct and confirm an analytically model for the isotropy degree $\rm iso(\gamma,M_A,\Delta v)=a+b\cos\gamma+c\cos^2\gamma+dM_A^{2/3}\cos\gamma +eM_A^{2/3}+f M_A^{2/3}\cos^2\gamma$. We perform numerical parameter studies to determine the coefficients. 
    \item We discuss the advantages of SFA in disentangling the galactic foreground and understanding the star formation.
\end{enumerate}

\section*{Acknowledgements}
Y.H. acknowledges the support of the NASA TCAN 144AAG1967. A.L. acknowledges the support of the NSF grant AST 1715754 and NASA ATP AAH7546. S.X. acknowledges the support for this work provided by NASA through the NASA Hubble Fellowship grant \# HST-HF2-51473.001-A awarded by the Space Telescope Science Institute, which is operated by the Association of Universities for Research in Astronomy, Incorporated, under NASA contract NAS5- 26555. 

\software{Julia \citep{2012arXiv1209.5145B}, ZEUS-MP/3D code \citep{2006ApJS..165..188H}, MATLAB}

\appendix
\section{Coefficients of the projected structure function}
\label{appendix.A}

As shown in \citet{2016MNRAS.461.1227K}, $D_z(\pmb{r})$ can be derived from the projection of structure function tensor for the velocity field:
\begin{equation}
    \begin{aligned}
      D_z(\pmb{r})=&2[(B(0)-B(r,\mu))+(C(0)-C(r,\mu))\cos^2\gamma-A(r,\mu)\cos^2\theta-2D(r,\mu)\cos\theta\cos\gamma]
    \end{aligned}
\end{equation}
where $\theta$ is the angle between the LOS and three-dimensional separation $\pmb{r}$. $\gamma$ is the inclination angle of the magnetic field relative to the LOS and $\mu(\gamma,\theta,\phi)=\sin\gamma\sin\theta\cos\phi+\cos\gamma\cos\theta$, where $\phi$ is the angle between the sky-projected $\pmb{r}$ and the POS magnetic field. To determine the coefficients $A(r,\mu)$, $B(r,\mu)$, $C(r,\mu)$, $D(r,\mu)$, we
firstly consider the velocity correlation tensor in the real space:
\begin{equation}
    \langle v_i(\pmb{x}_1)v_j(\pmb{x}_1+\pmb{r})\rangle=\int dkk^2d\Omega_ke^{i\pmb{k}\cdot\pmb{r}}A(k,\hat{\pmb{k}}\cdot\hat{\lambda})(\hat{\xi}_{\pmb{k}} \otimes \hat{\xi}_{\pmb{k}}^*)_{ij}
\end{equation}
here $\hat{\pmb{k}}$ is wavevector in Fourier space, $A(k,\hat{\pmb{k}}\cdot\hat{\lambda})$ is the power spectrum
, $\hat{\xi}$ is the direction of allowed displacement, and $\hat{\lambda}$ represents the mean direction of magnetic field. $(\hat{\xi}_{\pmb{k}} \otimes \hat{\xi}_{\pmb{k}}^*)_{ij}$ is a $\hat{\lambda}$ dependent tensor built from the displacement direction characteristic for the given mode.
\begin{figure*}
\centering
\includegraphics[width=1.0\linewidth,height=0.38\linewidth]{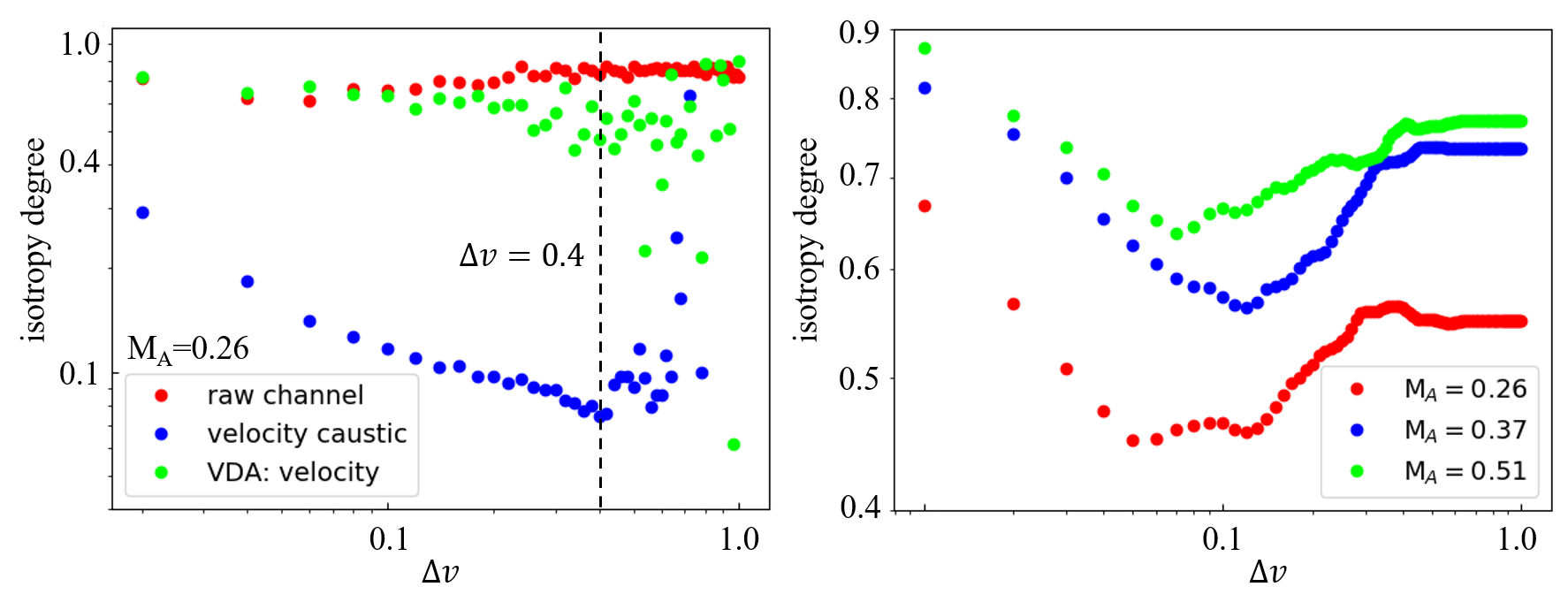}
\caption{\label{fig:sup_Ma}The correlation of isotropy degree with respect to velocity channel width $\Delta v$ and $\gamma=\pi/2$. Compressible simulations $\rm M_S\approx11.0$ are used here.}
\end{figure*}

In the case of Alfv\'en wave, the power spectrum  $A(k,\hat{\pmb{k}}\cdot\hat{\lambda})$ can be decomposed into spherical harmonics:
\begin{equation}
\begin{aligned}
A(k,\hat{\pmb{k}}\cdot\hat{\lambda})&=\sum_{l_1m_1}\frac{4\pi}{2l_1+1}A_{l_1}(k)Y_{l_1m_1}(\hat{\pmb{k}})Y_{l_1m_1}^*(\hat{\lambda})\propto k^{-11/3}\exp[-{\rm M_A}^{-4/3}\frac{|\mu_k|}{(1-\mu_k^2)^{2/3}}]\\
A_{l_1}(k)&=\frac{2l_1+1}{4\pi}\int\int A(k,\hat{\pmb{k}}\cdot\hat{\lambda})Y_{l_1m_1}^*(\hat{\pmb{k}})Y_{l_1m_1}(\hat{\lambda})d\Omega_kd\Omega_\lambda\\
T_{ll_1}(r)&=\int dk k^2j_l(kr)A_{l_1}(k)
\end{aligned}
\end{equation}
where $Y_{l_1m_1}$ is the spherical harmonics function, $\mu_k=\hat{\pmb{k}}\cdot\hat{\lambda}$, and $j_l$ is the spherical bessel function. The coefficients $A(r,\mu)$, $B(r,\mu)$, $C(r,\mu)$, and $D(r,\mu)$ then can be written as (see \citet{2016MNRAS.461.1227K} for details):
\begin{equation}
\label{eq.A4}
\begin{aligned}
A&=-8\pi\sum_{l,l_1,l_2}i^l(2l+1)(2l_2+1)\sqrt{\frac{(l-2)!(l_2-2)!}{(l+2)!(l_2+2)!}}T_{ll_1}
\times
\begin{pmatrix} 
	l & l_1 & l_2 \\
	0 & 0 & 0 \\
\end{pmatrix}
\begin{pmatrix} 
	l & l_1 & l_2 \\
	-2 & 0 & 2 \\
\end{pmatrix}
\frac{\partial^2P_l(\mu)}{\partial\mu^2},\\
B&=2\pi\sum_{l=0,2}i^lT_{ll}P_l(\mu)+4\pi\sum_{l,l_1,l_2}i^l(2l+1)(2l_2+1)\times
\sqrt{\frac{(l-2)!(l_2-2)!}{(l+2)!(l_2+2)!}}
\begin{pmatrix} 
	l & l_1 & l_2 \\
	0 & 0 & 0 \\
\end{pmatrix}
\begin{pmatrix} 
	l & l_1 & l_2 \\
	-2 & 0 & 2 \\
\end{pmatrix}
T_{ll_1}P_l^2(\mu),\\
C&=-2\pi\sum_{l=0,2}i^lT_{ll}P_l(\mu)-4\pi\sum_{l,l_1,l_2}i^l(2l+1)(2l_2+1)\times
\sqrt{\frac{(l-2)!(l_2-2)!}{(l+2)!(l_2+2)!}}
\begin{pmatrix} 
	l & l_1 & l_2 \\
	0 & 0 & 0 \\
\end{pmatrix}
\begin{pmatrix} 
	l & l_1 & l_2 \\
	-2 & 0 & 2 \\
\end{pmatrix}
\times
T_{ll_1}(2\mu^2\frac{\partial P_l(\mu)}{\partial \mu^2}+P_l^2(\mu)),\\
D&=8\pi\sum_{l,l_1,l_2}i^l(2l+1)(2l_2+1)\sqrt{\frac{(l-2)!(l_2-2)!}{(l+2)!(l_2+2)!}}T_{ll_1}
\times
\begin{pmatrix} 
	l & l_1 & l_2 \\
	0 & 0 & 0 \\
\end{pmatrix}
\begin{pmatrix} 
	l & l_1 & l_2 \\
	-2 & 0 & 2 \\
\end{pmatrix}
\mu\frac{\partial^2P_l(\mu)}{\partial\mu^2},\\
\end{aligned}\\
\end{equation}
where $P_l(\mu)$ is the Legendre polynomial.
\begin{equation} 
\begin{pmatrix} 
	l & l_1 & l_2 \\
	0 & 0 & 0 \\
\end{pmatrix}
and 
\begin{pmatrix} 
	l & l_1 & l_2 \\
	-2 & 0 & 2 \\
\end{pmatrix}
\end{equation}
are Wigner's 3-j symbols. By performing Legendre expansion for the coefficients $A(r,\mu)$, $B(r,\mu)$, $C(r,\mu)$, $D(r,\mu)$ up to the second order, i.e., $A(r,\mu)=\sum_nA_n(r)P_n(\mu)\approx A_0(r)+A_2(r)P_2(\mu)$, $D_z(\pmb{r})$ can be further simplified to:
\begin{equation}
\begin{aligned}
D_z(\pmb{r})&\approx c_1-c_2\cos\phi-c_3\cos^2\phi\\
\end{aligned}
\end{equation}.

The coefficients are expressed as:

\begin{equation}
\begin{aligned}
c_1 &= (q_1+q_2\cos^2\theta+q_3\cos^4\theta)r^\nu\\
c_2 &= (s_1+s_2\cos^2\theta)r^\nu\sin\theta\cos\theta\sin\gamma\cos\gamma\\
c_3 &= (u_1+u_2\cos^2\theta)r^\nu\sin^2\theta\sin^2\gamma\\
q_1 &= 2(B_0(0)-B_0)+2(C_0(0)-C_0)\cos^2\gamma+B_2+C_2\cos^2\gamma\\
q_2 &= -2A_0+A_2-4D_1\cos\gamma-3(B_2+C_2\cos^2\gamma)\cos^2\gamma\\
q_3 &= -3A_2\cos^2\gamma\\
s_1 &= 6(B_2+C_2\cos^2\gamma)+4D_1\\
s_2 &= 6A_2\\
\mu_1 &= 3(B_2+C_2\cos^2\gamma)\\
\mu_2 &= 3A_2\\
\end{aligned}
\end{equation}
Note Eq.~\eqref{eq.A4} only gives the coefficients for Alfv\'en wave. The coefficients for fast and slow wave can be found in  \citet{2016MNRAS.461.1227K}.

\section{Supersonic turbulence}
\label{Appendix.B}

The density effect in subsonic turbulence is insignificant since the density statistics passively follow the velocity statistics \citep{2005ApJ...624L..93B}. As shown in Fig.~\ref{fig:iso_vda}, density has little effect for channel width $\Delta v\le0.3$. Also, density contribution in thin channel can be removed by VDA. However, density fluctuation in supersonic turbulence has different properties, which can change the anisotropy. In Fig.~\ref{fig:sup_Ma}, we investigate the density effect in the supersonic simulation $\rm M_S=10.81$ and $\rm M_A=0.26$. We plot the correlation of isotropy degree with respect to velocity channel width $\Delta v$ at $\gamma=\pi/2$. We find the isotropy degree for the raw channel map (i.e., real density field) appears almost like a flat curve, showing isotropy degree $\approx0.7$. The pure velocity caustic map (i.e., uniform density field) exhibits a steep isotropy curve until $\Delta v\approx0.4$. It indicates that the flatten curve for the real density field case comes from density's contribution. To remove the density effect, several experiments have been test in \citet{2020arXiv201215776Y} using the VDA technique. Here we briefly describe one of the possibilities.

To extract the velocity information, one can use the low-density sampling method. This method selects the data points whose corresponding column density is low to calculate the structure-function. For instance, \citet{XH21} use only 10,000 data points with the lowest column density to calculate the structure-function of velocity centroid. They successfully retrieve velocity's anisotropy in this way. This low-density sampling is based on the fact that low-density statistics still exhibit scale-dependent anisotropy arise from shearing by Alfv\'en waves \citep{2005ApJ...624L..93B} and the SFA requires only a few valid data points to retrieve the anisotropy \citep{SFA}. 

We test this low-density sampling method in Fig.~\ref{fig:sup_Ma}. We select the data points for each channel map whose corresponding column density is lower than the $10^{th}$ percentile threshold. The threshold results in approximately 65,000 valid data points. We set $\gamma=\pi/2$ and $R=10$ pixels. As shown in Fig.~\ref{fig:sup_Ma}, the isotropy degree gets retrieved. It decreases until $\Delta v\approx0.1$, after which the density contribution gets dominated again. Therefore, in supersonic turbulence, one can still get rid of the density contribution by selecting only low-density data points in a thin channel. Comparing with the pure velocity caustic case, the isotropy degree calculated from low-density  data points is overestimated. The coefficients for determining $\gamma$ in supersonic case might be different. Nevertheless, the coefficients can be established empirically from numerical simulations.

\section{Anisotropies of Alfv\'en, fast, and slow modes}
\label{Appendix:C}
\begin{figure*}
\centering
\includegraphics[width=1.0\linewidth,height=0.63\linewidth]{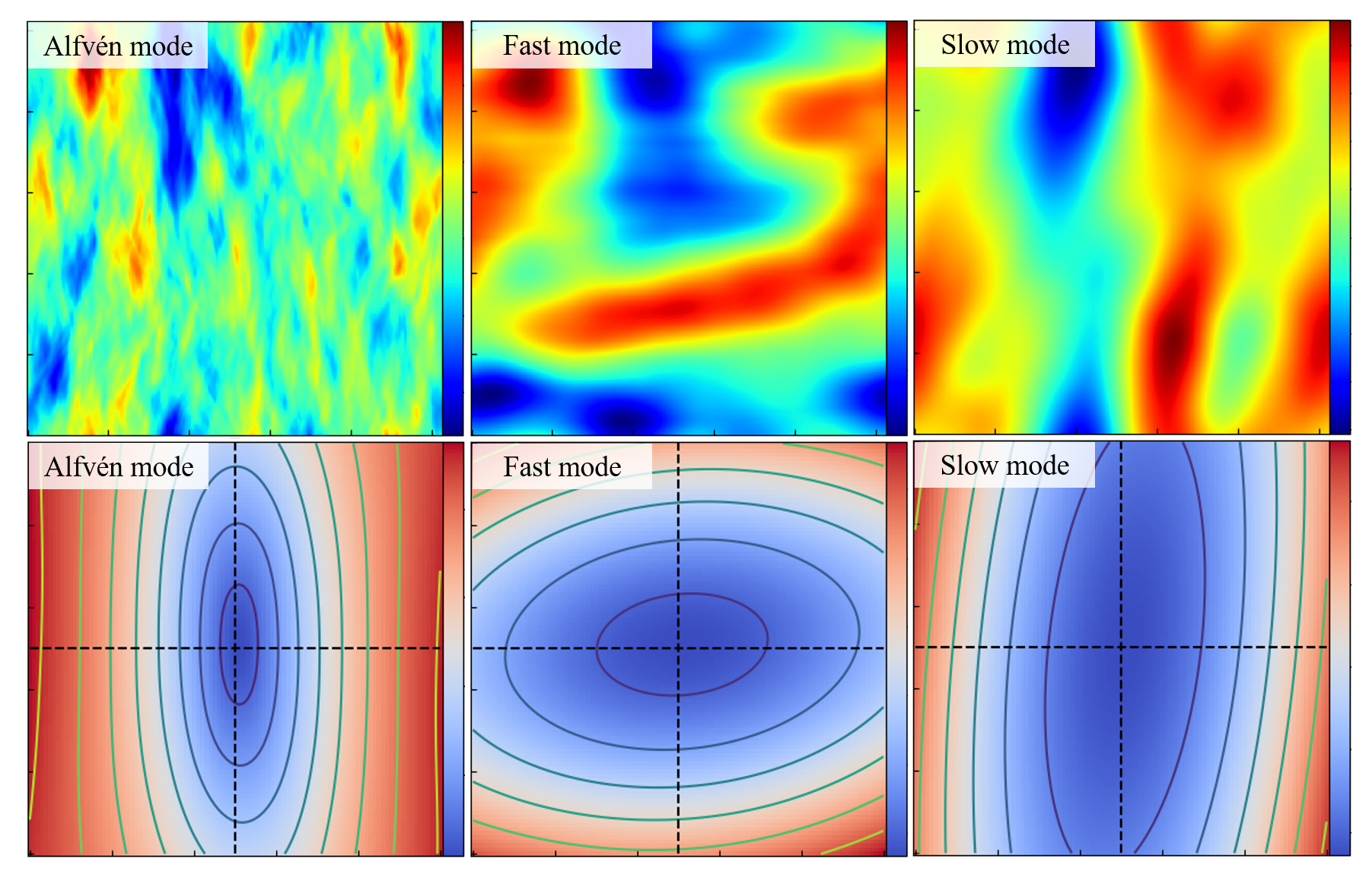}
\caption{\label{fig:mode}\textbf{Top:} Projected velocity maps  of Alfv\'en (left), fast (middle), and slow (right) modes in conditions of $\rm M_A=0.52$ and $\rm M_S=2.17$. The mean magnetic field is along the vertical direction. \textbf{Bottom:} Projected structure-functions of Alfv\'en (left), fast (middle), and slow (right) velocity modes. The structure-functions are calculated in the global reference frame. All plots are on the scale of 50 pixels from center to boundary.}
\end{figure*}

Compressible MHD turbulence consists three MHD modes, i.e., incompressible Alfv\'en mode, compressible fast mode, and compressible slow mode. To investigate the anisotropies of different modes, we employ the mode decomposition method proposed in \citet{2003MNRAS.345..325C}. The decomposition is performed in Fourier space by projecting the velocity’s Fourier components on the direction of the displacement vectors $\hat{\zeta_A}$, $\hat{\zeta_f}$, and $\hat{\zeta_s}$ for Alfv\'en, fast, and slow modes, respectively. The displacement vectors are defined as:
\begin{equation}
    \begin{aligned}
    \hat{\zeta_A}&\propto\hat{k}_\bot\times\hat{k}_\parallel\\
    \hat{\zeta_f}&\propto(1+\beta/2+\sqrt{D})k_\bot\hat{k}_\bot+(-1+\beta/2+\sqrt{D})k_\parallel\hat{k}_\parallel\\
    \hat{\zeta_s}&\propto(1+\beta/2-\sqrt{D})k_\bot\hat{k}_\bot+(-1+\beta/2-\sqrt{D})k_\parallel\hat{k}_\parallel\\
    \end{aligned}
\end{equation}
where wavevectors $\vec{k}_\parallel$ and $\vec{k}_\bot$ are the parallel and perpendicular to the mean magnetic field $\vec{B}_0$, respectively. $D = (1 + \beta/2)^2 - 2\beta\cos^2 \vartheta$, $\beta=2(M_A/M_S)^2$, and $\cos \vartheta=\hat{k}_\parallel\cdot\hat{B}_0$. 

We only decompose the LOS velocity component and calculate the second-order structure function in the global reference frame. The projected (along the LOS) 2D velocity maps and structure functions are plotted in Fig.~\ref{fig:mode}. The contours, i.e., the anisotropies, of Alfv\'en and slow modes are elongating along the vertical direction, which is also the direction of the mean magnetic field. Fast mode, however, is isotropic only in the zeros approximation. It can still exhibit anisotropy perpendicular to the Alfv\'en and slow modes. Nevertheless, Alfv\'en mode is the most important component of MHD turbulence \citep{2003MNRAS.345..325C,2007mhet.book...85S} and contributes the most significant anisotropy.  Therefore, we expect the fitting function Eq.~\ref{eq.iso} is applicable to the majority of ISM environments, as shown in our numerical experiments (see \S~\ref{sec:results}).

%\bibliography{sample63}{}
%\bibliographystyle{aasjournal}



\end{document}